# Improved Calculation of the Young's Modulus of Rectangular Prisms from their Resonant Frequency Overtones by Identifying Appropriate Shear Constants

Paul A. Bosomworth[1], Rui Zhang[2, &], Lawrence M. Anovitz[2, *]

[1]BuzzMac International LLC, Portland, ME 04101, United States
[2]Chemical Sciences Division, Oak Ridge National Laboratory, Oak Ridge, TN 37831, United States

[*]Corresponding author: Lawrence M. Anovitz, anovitzlm@ornl.gov
[&]Present address: Eyring Materials Center, Arizona State University, Tempe, AZ 85287, United States

**Abstract**

Young's modulus is an important parameter for characterizing the strength of, and wave propagation through, a given material. This study improves the estimation of Young's modulus using the impulse excitation (IE) technique based on an experimental analysis of 19 borosilicate glass bars. Analysis of the frequency equations relating Young's modulus to the out-of-plane and in-plane flexural resonant frequencies of rectangular prisms has been conducted for both the fundamental frequency and its overtones at higher orders of vibration. The Young's modulus of three novaculite rocks with various porosities were then measured up to the seventh order of vibration to validate the optimum shear constant equation for estimating Young's modulus. Young's modulus was found to be nearly frequency independent for these rock samples.

**Keywords:** Mechanical properties; Impulse excitation; Silica glass; Novaculite

# 1   Introduction

## 1.1   Measurement of elastic properties

Mechanical properties are important parameters in various areas of science and technology including material sciences, civil engineering, and geosciences. The elastic properties of a material can be measured using either static or dynamic testing methods. Dynamic methods such as the impulse excitation (IE) technique and resonant ultrasound spectroscopy (RUS) have been shown to be more accurate and repeatable than static ones such as four-point bending and nanoindentation [1], at least for isotropic materials. The advantages of dynamic methods include the relatively low stresses involved during the measurement and the fact that the sample remains undamaged after testing. In addition, damping properties such as internal friction can also be measured that can be related to porosity, grain boundary phases and microstructural features of a given material [2].

The IE technique is perhaps the simplest and the most convenient dynamic method available. A wide range of sample sizes and shapes can be tested using relatively low-cost equipment. In this method, a sample of uniform cross section (typically a rectangular prism, cylinder, or disc) is excited mechanically by striking it with a small hammer (i.e., impulse tool). The sample is supported and struck in such a way that it predominantly vibrates in only one mode. By knowing the vibration mode and the sample's shape, size, and density, the elastic properties can be calculated from the measured resonant frequencies, which include the fundamental frequencies and their overtones. If the fundamental resonant frequency is used, one can expect errors in the calculated Young's and shear moduli to be $\leq 1$ % for rectangular prisms with length/width ratios > 2 [3].

The present study is part of a series designed to reveal the relationship between changes in elastic properties and microstructure in geologic materials. However, it was unknown if some of these elastic properties were also dependent on the strain rate. Because of this uncertainty, the Young's and shear moduli needed to be calculated not just from the fundamental frequencies but also from their overtones using the appropriate modes of vibration. Although explicit frequency equations relating the shear modulus to the fundamental torsional frequency and overtones are available [4-7], the calculations have been shown to be imprecise [2], and the case for Young's modulus derived from the flexural mode of vibration is more complicated. In general, no explicit approximate solutions are available for vibration orders > 2. In this case, solutions require a shape factor, or so-called shear constant, whose value has not been established unequivocally. This paper addresses this issue.

## 1.2 Free flexural vibrations of a Timoshenko beam

The flexural mode of vibration is the most often used to obtain the Young's modulus using the IE technique. The longitudinal mode can also be used, but the fundamental resonant frequency is much higher, as are its overtones. Thus, it's much more difficult to excite higher orders of vibration. For the flexural vibration mode, the sample can be considered as a Timoshenko beam. In a typical deformed beam, the shear stresses are highest around the neutral axis. This is, consequently, where the largest shear deformation takes place. Because of this, during deformation the cross-section curves. The Timoshenko beam theory, however, retains the assumption that the cross-section remains planar during bending. Under this assumption, the governing equation is [8]:

$$EI\frac{\partial^4 y}{\partial x^4} + m\frac{\partial^2 y}{\partial t^2} - \rho I\left(1 + \frac{E}{k'G}\right)\frac{\partial^4 y}{\partial x^2 \partial t^2} + \frac{\rho^2 I}{k'G}\frac{\partial^4 y}{\partial t^4} = 0 \tag{1}$$

where $E$ is Young's modulus; $I$ is the second moment of area with respect to the centroidal z-axis, normal to the vibration direction; $y$ is the displacement in the vibration direction; $x$ is the coordinate in the direction of length; $t$ is time; $\rho$ is the mass density; $m = \rho A$, where $A$ = cross-sectional area; $k'$ is the shear constant, introduced by Timoshenko to account for the effect of shear on the slope of the elastic line, defined as the ratio of average unit shear across a section to the unit shear at the neutral axis; and $G$ is the shear modulus. It is important to note that, in the rest of this paper, $t$ = thickness, not time.

We can see from eq. 1 that the third and fourth terms added by Timoshenko [9] account for shear and rotary inertia, which depend on $G$ and $k'$. For a beam in which both ends are freely vibrating the bending moment and shearing force are zero at these ends. In practice, this requires the sample to be supported at two vibrational nodes. Under free boundary conditions, the following equation can be used [10]:

$$2 - 2\cos h(\alpha)\cos(\beta) + \frac{\gamma^2\left(\gamma^4 r^2(r^2-s^2)^2+(3r^2-s^2)\right)}{\sqrt{1-\gamma^4 r^2 s^2}}\sinh(\alpha)\sin(\beta) = 0 \qquad (2)$$

where $r^2 = \frac{I}{Al^2}$, $s^2 = \frac{EI}{k'GAl^2}$, $\gamma^4 = \frac{\rho Al^4}{EI}\omega^2$, $l$ = length, $\omega = 2\pi f$ where $f$ = frequency and $\alpha$ and $\beta$ are eigenvalues, given by:

$$\alpha_n = \frac{\gamma^2}{\sqrt{2}}\sqrt{-(r^2+s^2)+\sqrt{(r^2-s^2)^2+\frac{4}{\gamma^4}}} \text{ and } \beta_n = \frac{\gamma^2}{\sqrt{2}}\sqrt{(r^2+s^2)+\sqrt{(r^2-s^2)^2+\frac{4}{\gamma^4}}} \qquad (3)$$

The frequency is then given by:

$$f_n = \frac{\gamma_n^2}{2\pi}\sqrt{\frac{EI}{\rho Al^4}} \qquad (4)$$

where $n$ = order of vibration (1=fundamental, 2=1st overtone, etc.). Substituting $k^2$ for $I/A$ where $k$ = radius of gyration, $E$ can be expressed as:

$$E = \left(\frac{2\pi f_n l^2}{k\gamma_n^2}\right)^2 \rho \qquad (5)$$

Note: $k = \frac{t}{\sqrt{12}}$ for a rectangular prism. To solve the eigenvalues in this study, Mathematica software from Wolfram Research, Inc was used. A significant factor affecting the accuracy of the calculations is the assumed value of $k'$. Several values and equations have been proposed, with the most common ones depending on Poisson's ratio ($\mu$), as shown in Table 1.

For the fundamental frequency (n=1), $E$ was calculated using well-established equations based on the works of Goens [11], Pickett [12], and Spinner et al., [4, 13], which form the basis of the equations cited in the ASTM standards for the IE technique such as E1876-22 and C1259-21. E can be expressed in a similar manner to eq. 5 as:

$$E = \left(\frac{2\pi f_1 l^2}{k\, m_1^2}\right)^2 \rho T_1 \qquad (6)$$

where $m_1$ = first solution for $\cos(m)\cosh(m) = 1$ which is $\cong 4.73004$, and

$$T_1 = 1 + 79.02 \left(1 + 0.0752\mu + 0.8109\mu^2\right) \left(\frac{k}{l}\right)^2 - 125 \left(\frac{k}{l}\right)^4 -$$

$$\left[\frac{1201 \left(1 + 0.2023\mu + 2.173\mu^2\right) \left(\frac{k}{l}\right)^4}{1.000 + 76.06 \left(1 + 0.1408\mu + 1.536\mu^2\right) \left(\frac{k}{l}\right)^2}\right] \tag{7}$$

However, the correction factor $T_1$ and therefore $E$ are dependent on Poisson's ratio. For a linearly elastic, isotropic material, the Poisson's ratio can be calculated from $E$ and $G$ as:

$$\mu = \frac{E}{2G} - 1 \tag{8}$$

Therefore, an iterative method is required to calculate $E$ and $\mu$ as described in the ASTM standards.

## 2    Materials and methods

### 2.1    Materials

Glass was chosen as the test-bar material for this study because it has no internal grain boundaries and limited defects, and it can, therefore, be reasonably assumed that the Young's and shear moduli are independent of the order of resonant frequency (n) for a given mode of vibration. Soda-lime silica glass bars were obtained from Swift Glass, and samples with a wide range of width/length and width/thickness ratios were prepared from the same original glass plate. The physical properties of these bars are shown in Table 2. Note that the density increase with width may be because uncertainties in the dimensional measurements due to a combination of sample machining tolerances, micrometer precision (0.001 mm), and the measurement method, in which the vertical sides of the samples may not have been perfectly parallel. In addition, three grades of novaculite: soft Arkansas (Medium), hard Arkansas (Fine), and translucent Arkansas (Extra Fine) were used to test and validate the proposed Young's modulus solutions. These were purchased from Dan's Whetstones. The physical properties of the prepared novaculites are shown in Table 3.

## 2.2    Impulse excitation testing

A commercial non-destructive testing (NDT) instrument, Buzz-o-sonic (BuzzMac International, LLC), was used to measure the resonant frequencies of the rectangular prisms using the impulse excitation technique as described in ASTM E1876-22. The resonant frequencies were recorded with an integrated electronics piezoelectric microphone (1/4" free-field, prepolarized 377C01 microphone and 426B03 preamplifier, PCB Piezotronics, Inc.) connected to a signal

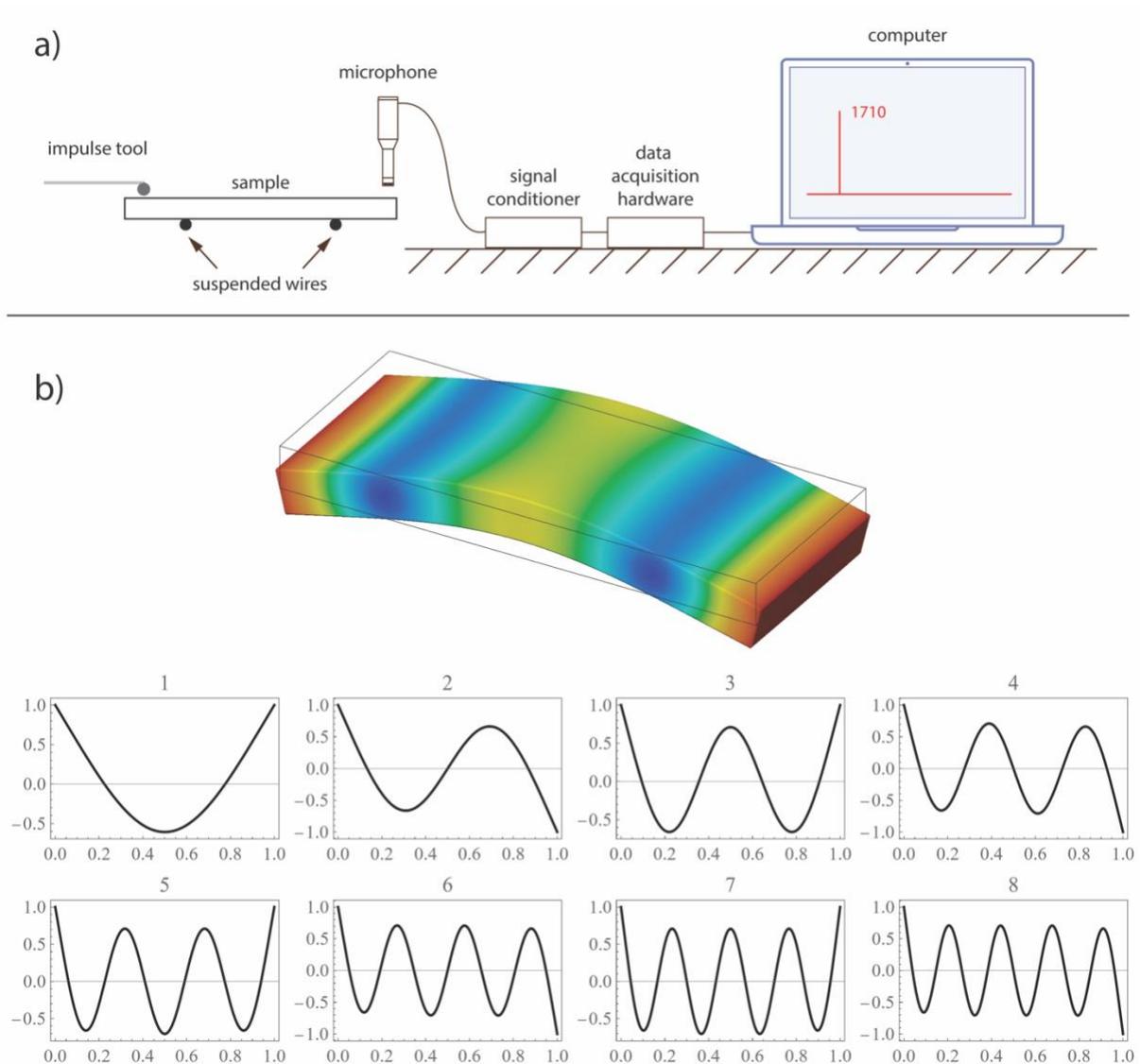

**Figure 1** (a) Impulse excitation setup [2] and (b) schematic of out-of-plane flexural mode and an example for the flexural modes (y axes = normalized amplitude, x axes = normalized length).

conditioner (480B21, PCB Piezotronics, Inc.), the output of which was sent to data acquisition hardware (500 kHz sample rate, 16-bit resolution, NI 9222, NI) attached to a computer running the Buzz-o-sonic software. The samples were excited using an impulse tool as per ASTM E1876-22 (~5 mm ball bearing cemented to a ~100 mm flexible polymer rod). The setup is shown in Figure 1a.

The samples were tested in flexure up to the seventh order of vibration. There are two flexural modes:

1. Out-of-plane (*flatwise*): a flexure mode in which the direction of displacement is normal to the major plane of the test specimen.

2. In-plane (*edgewise*): a flexure mode in which the direction of displacement is in the major plane of the test specimen.

The samples were freely supported on the vibrational nodes using piano steel wires of 0.229 mm (0.009") diameter. Parallel wires were used in all cases. The wires were placed equidistant from each end of the sample. For example, for the fundamental flexural frequency, the wires were placed at the nodes located at 0.224 times the length of the sample from each end of the sample. Where possible, the wires were tried at alternative nodal positions to increase the frequency response for a given mode and order of vibration. This was often the case for $n \geq 5$. The microphone was placed over the antinodes and several alternative positions were tried. An example of the fundamental vibration and seven overtones for the out-of-plane flexural (bending) mode is shown in Figure 1b.

### 3    Results and discussion

#### 3.1    Selection of shear constants using the glass bars

The appropriate or optimum shear constant, $k'_{opt}$, for a given mode of vibration for each glass bar was determined as follows:

1.  The shear constant was varied over the range indicated in Table 1 (~2/3 – 1.3) in 0.001 increments. The upper range was chosen so that high values from the Hutchinson model would be included.

2.  For each shear constant, eq. 5 was used to determine the Young's modulus for each order of vibration.

3.  The sample standard deviation of these Young's moduli was obtained.

4.  The shear constant giving the minimum standard deviation was chosen as the optimum value.

To calculate the Young's moduli using eq. 5, the out-of-plane flexural frequencies and shear moduli of the glass bars were required. The measured frequencies and estimated shear moduli are shown in Tables 4 and 5. The latter were obtained using the approach described in part 1 of this work [2]. The shear modulus derived from the fundamental torsional frequency was used rather than that measured at each overtone because this value was expected to have less uncertainty due to the correction factors involved [2].

The standard deviations in the Young's moduli (out-of-plane flexure) as a function of the value of the shear constant are shown for selected glass bars in Figure 2 and for all the glass bars in Figure S1 in the Supporting Information (SI). Note that the width/thickness (b/t) ratio of the bars increases with increasing sample number (see Table 2). For comparison purposes, the various shear constants suggested in the literature are indicated. For the shear constants that depend on

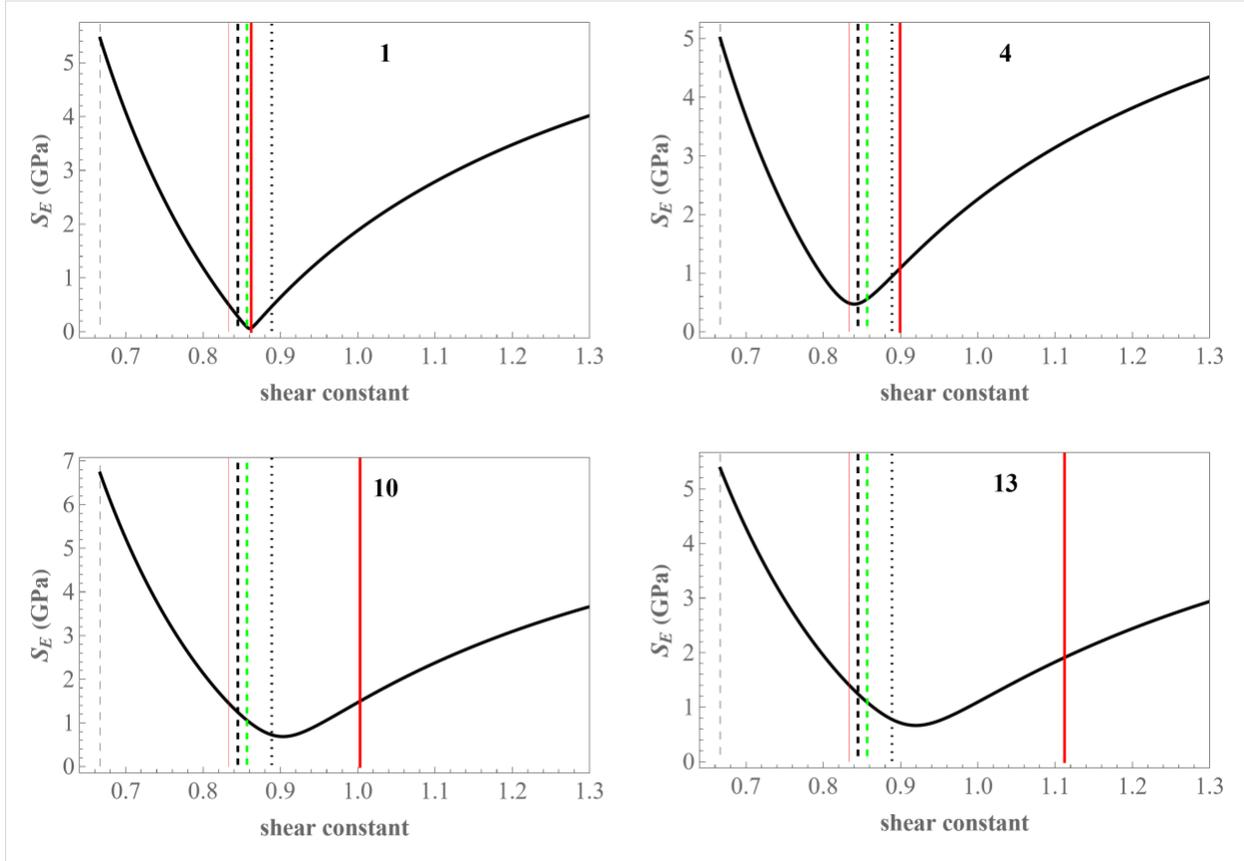

**Figure 2** Standard deviation of Young's modulus ($S_E$) as a function of the shear constant, showing the optimum shear constants for the selected glass bars (out-of-plane flexural vibrations, n=1-7). Timoshenko (thin gray dashes), Ritchie (thin red line), Picket (black dots), Cowper (black dashes), Kaneko (green dashes), Hutchinson (thick red line).

Poisson's ratio, a mean value of 0.197 derived from all the glass samples was used (Table 5). For the case of Hutchinson's shear constant, Poisson's ratio was calculated from the values of E and G using eqs. 5 and 8 using an iterative method in Mathematica. An example of this calculation is shown in the Appendix. The root-mean-square (RMS) values for the differences between the average optimum shear constant of all the glass samples (n=1) and the literature shear constants (Table 1) are shown in Table 6.

Comparing the standard deviation curves with the proposed shear constants shows, first, that the Timoshenko shear constant deviated the most from the optimum value at the lowest the length to thickness (b/t) ratio, and second, that the deviation of Hutchinson's shear constant from the optimum Young's moduli from eq. 5 increased as the width to thickness ratio (b/t) increased (increasing sample number). Consequently, to obtain the optimum shear constants for lower b/t ratios, the in-plane flexural frequencies were also measured on several samples (samples 1, 12, 18, and 19, Table 7). The standard deviations in Young's modulus (in-plane) vs. shear constant are shown for these glass bars in Figure 3. Since some of the in-plane flexural frequencies could not be excited reliably at n > 5 (samples 1 and 18), the standard deviation was taken over the order of frequencies n=1-5. Once again, for the shear constants depending on Poisson's ratio, a mean value of 0.197 was used (Table 5); and for Hutchinson's shear constant, the Poisson's ratio was calculated from the values of E and G using eqs. 5 and 8 and an iterative method in Mathematica.

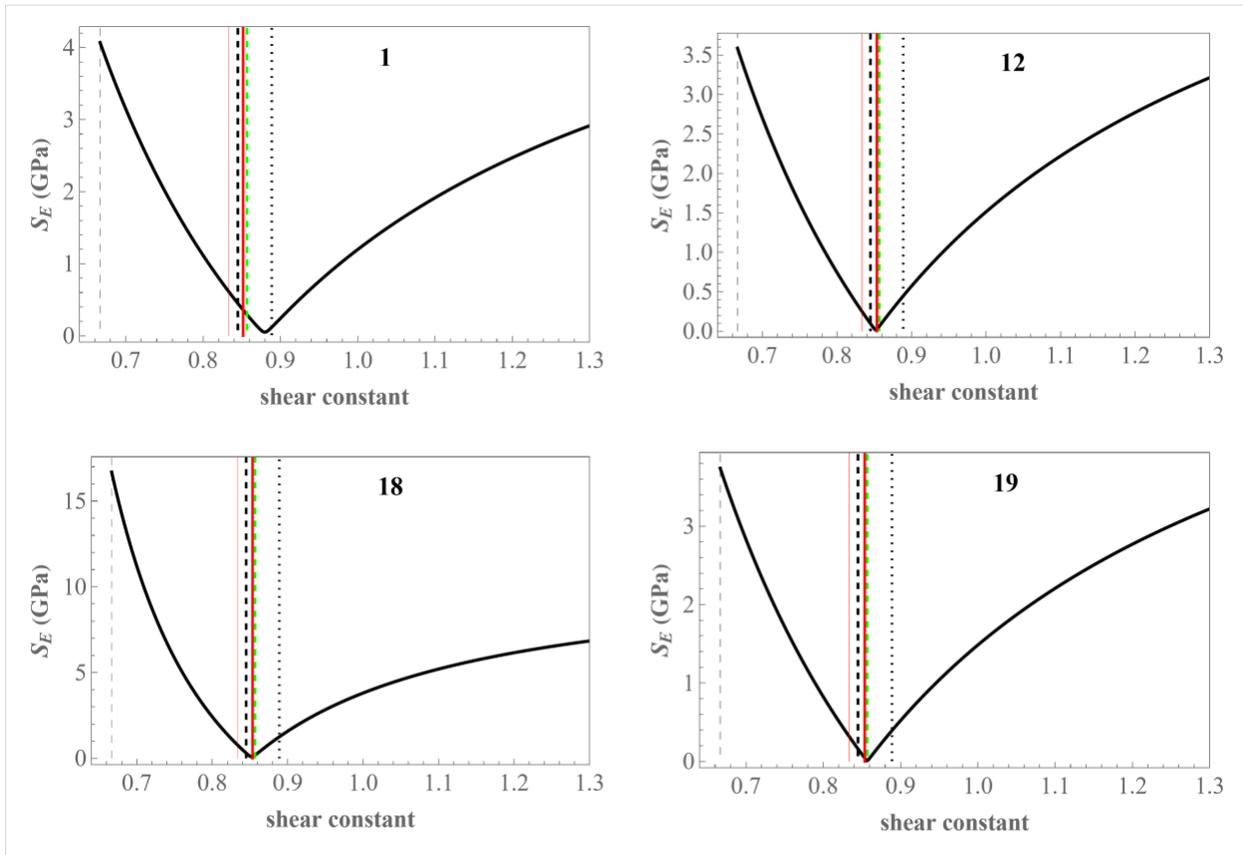

**Figure 3** Standard deviation of Young's modulus ($S_E$) as a function of the shear constant showing the optimum shear constants of the glass bars (in-plane flexural vibrations, n=1-5). Timoshenko (thin gray dashes), Ritchie (thin red line), Picket (black dots), Cowper (black dashes), Kaneko (green dashes), Hutchinson (thick red line).

The RMS values of the differences between the average optimum shear constant of all the glass samples (n=1) for the in-plane measurements and the literature shear constants are shown in Table 8.

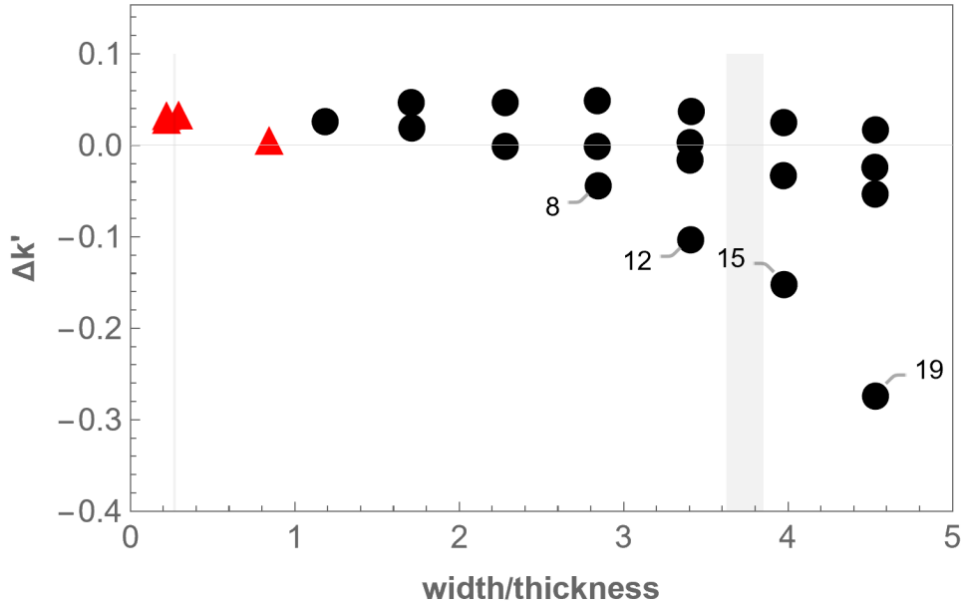

**Figure 4** Difference between Pickett's shear constant and the optimum ($\Delta k'$). Out-of-plane flexure mode (black circles), in-plane flexure mode (red triangles). Similar results were obtained with Kaneko and Cowper's shear constants.

The above RMS results show that the best literature shear constants for out-of-plane flexure are those of Pickett, Kaneko, and Cowper (Table 6) except for the longer samples (samples 8, 12, 15, and 19, Table 2). For in-plane flexure the best constants are those of Kaneko, Hutchinson, and Cowper (Table 8). Hutchinson's shear constant deviated from the optimum at higher width/thickness (b/t) ratios (> 2.4) except for the longer samples where b/l ~ 0.1. The differences between the Pickett and Hutchinson shear constants and the optimum values are shown in Figures 4 and 5, respectively. Similar results to Pickett (Figure 4) were obtained with the Cowper and Kaneko shear constants (Figure S2). The shaded areas in Figures 4 and 5 indicate the width/thickness ratio of the rock samples.

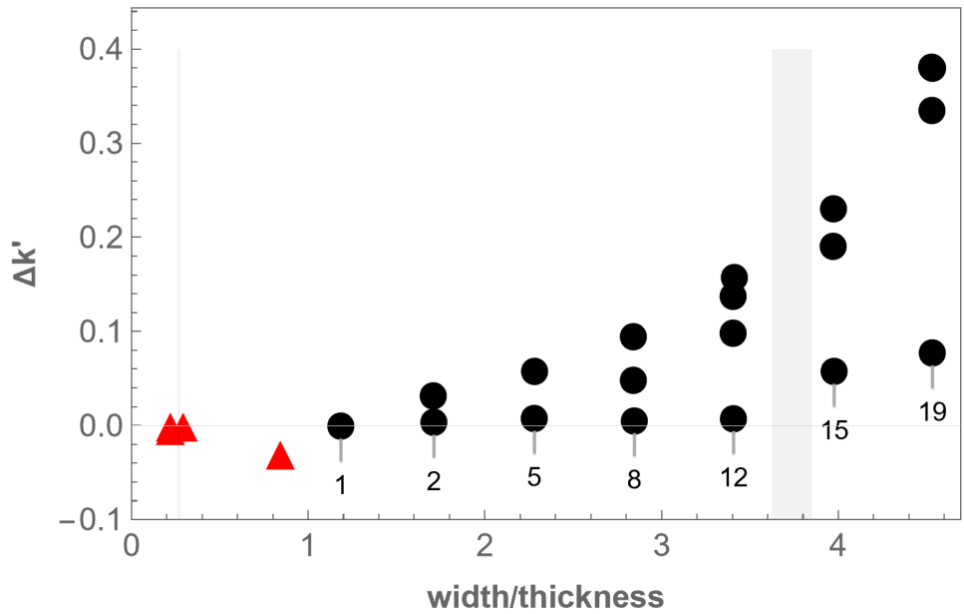

**Figure 5** Difference between Hutchinson's shear constant and the optimum ($\Delta k'$) as a function of the width to thickness ratio of the samples. Out-of-plane flexure mode (black circles), in-plane flexure mode (red triangles).

The dependence of $\Delta k'$(Hutchinson) on the width/thickness and length/thickness for the glass bars is shown in Figure S3. The samples with some of the lowest deviations are labeled in Figure S3. These have similar width/length ratios of ~0.1 (Figure 6).

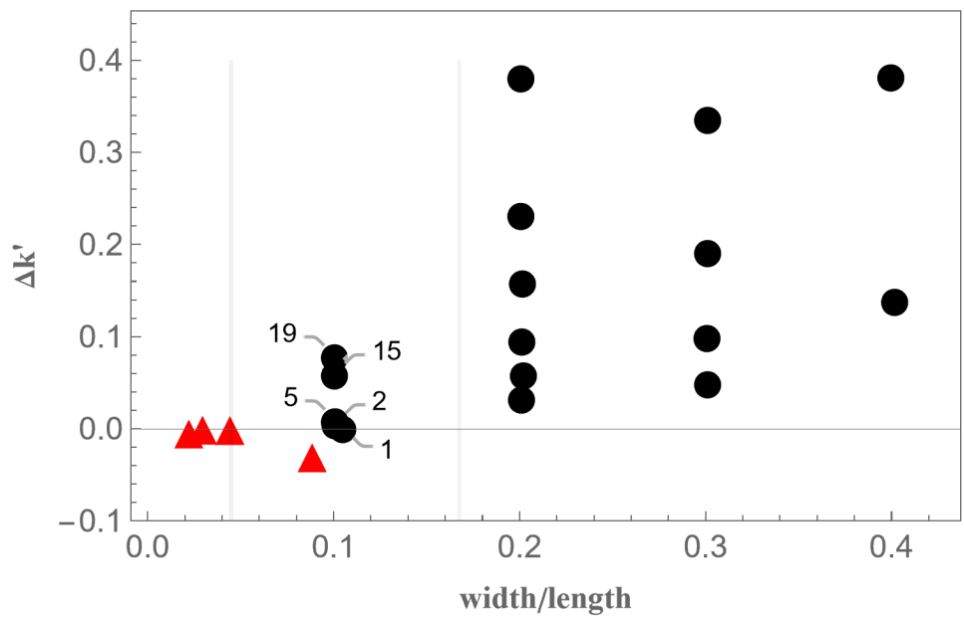

**Figure 6** Difference between Hutchinson's shear constant and the optimum ($\Delta k'$) as a function of the width to length ratio of the samples. Out-of-plane flexure mode (black circles), in-plane flexure mode (red triangles).

The mean Young's moduli calculated using eq. 5 and the Pickett, Cowper, Kaneko, and Hutchinson shear constants for each order of vibration for all the glass samples are shown in Figures 7 and 8. These results shows that the Young's moduli derived from the out-of-plane flexural mode are higher than those derived from the in-plane flexural mode (Figure 9), and that Hutchinson's shear constant gives the smallest difference between these two modes.

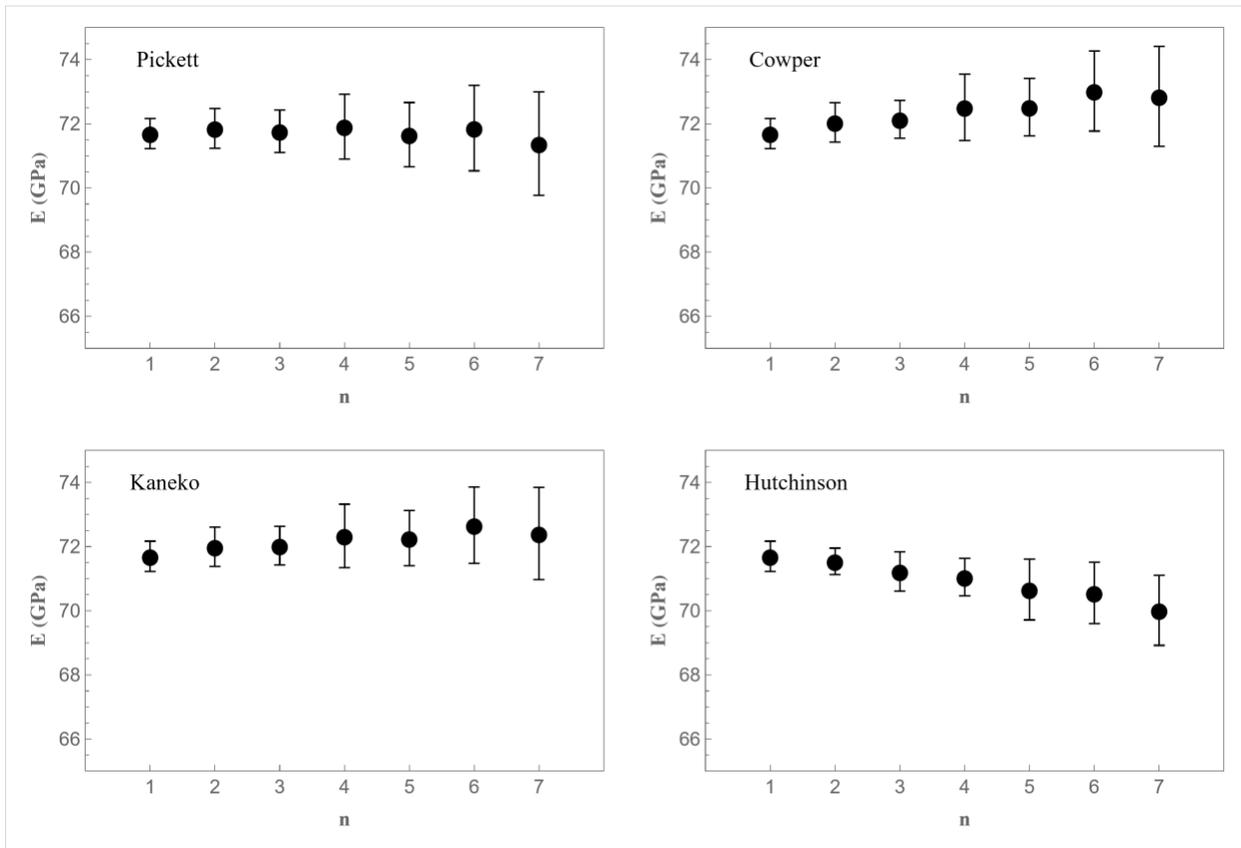

**Figure 7** Mean Young's modulus (out-of-plane flexural mode) vs. order of vibration for all the 19 glass bars. Error bars = ± standard deviation.

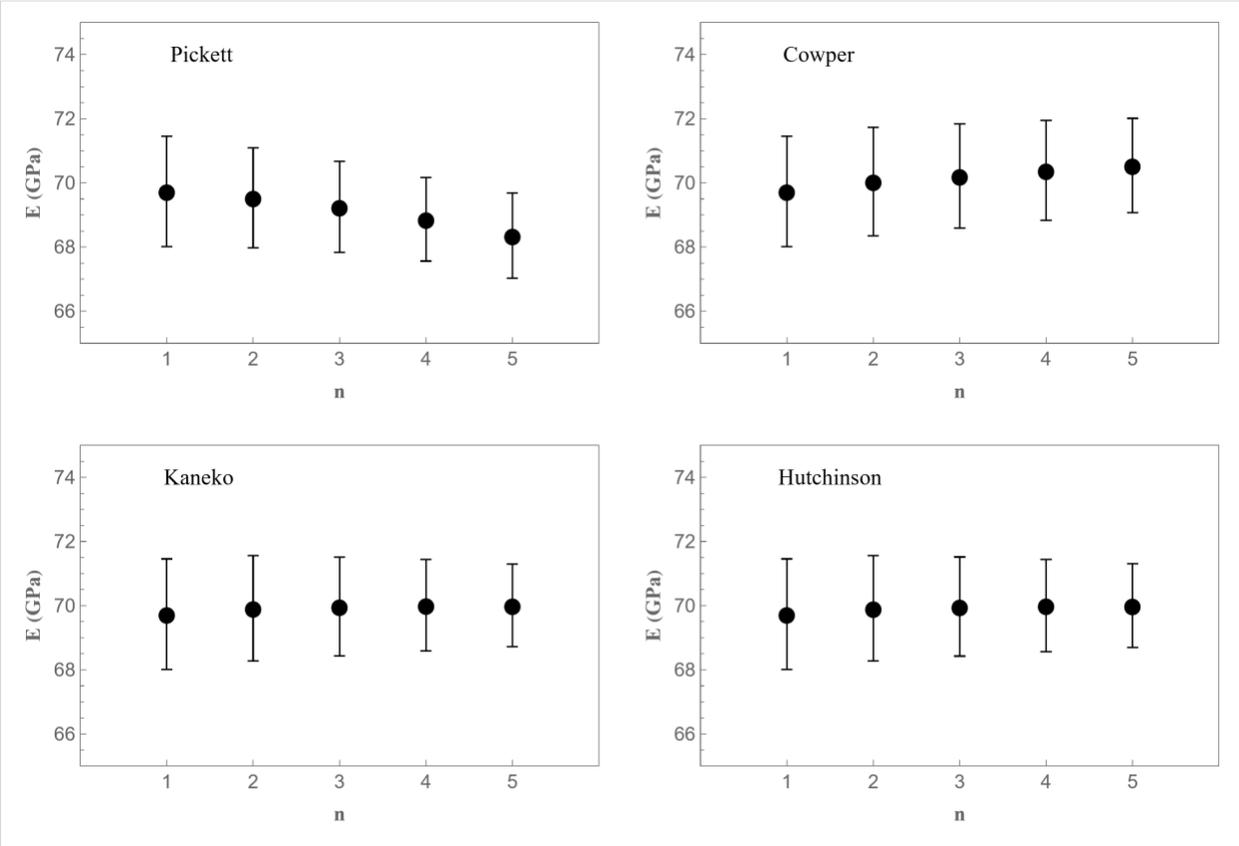

**Figure 8** Mean Young's modulus (in-plane flexural mode) vs. order of vibration for the glass bars of samples 1, 12, 18, and 19. Error bars = ± standard deviation.

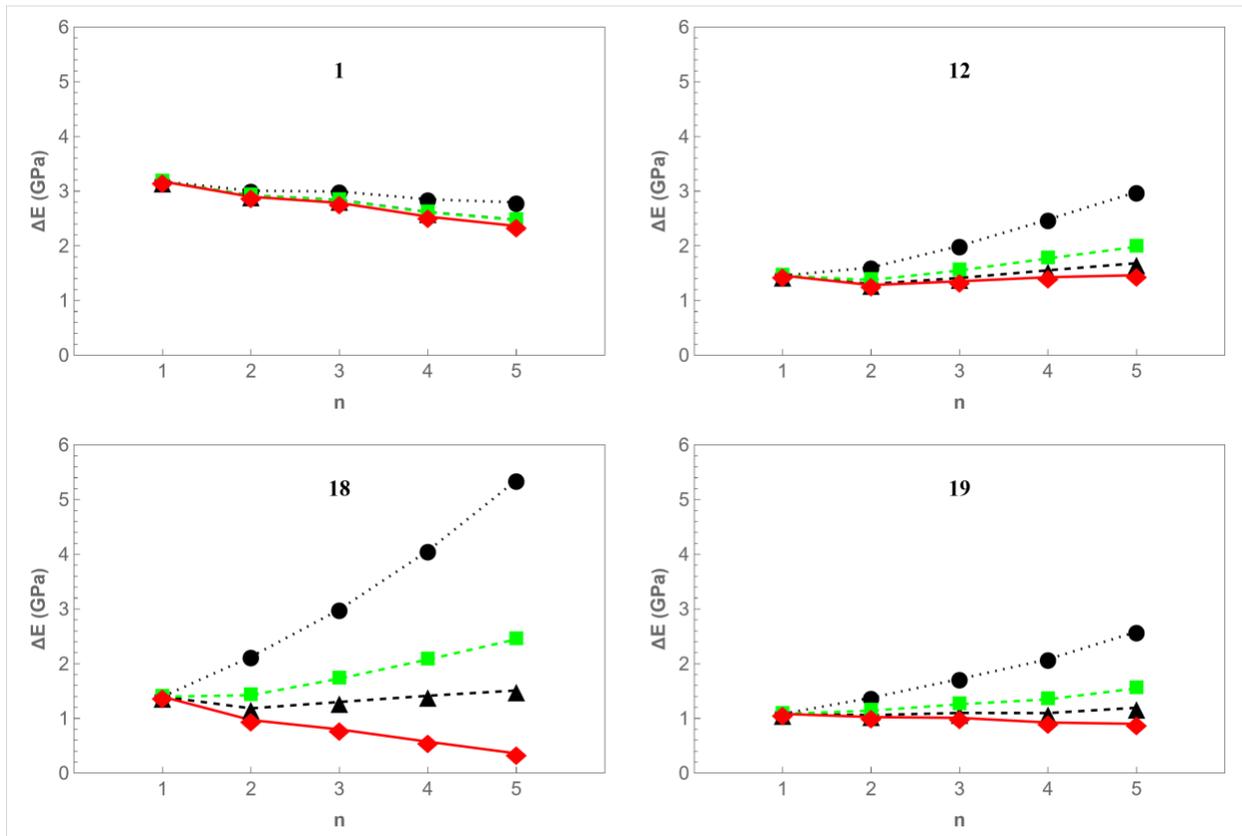

**Figure 9** The difference in Young's moduli from out-of-plane and in-plane flexural modes of vibration. Pickett (black circles and dots), Cowper (green squares and dashes), Kaneko (black triangles and dashes), Hutchinson (red diamonds and lines).

### 3.2 Validation of the shear constants using the novaculites

The shear moduli of the novaculites determined in part 1 of this study [2] are shown in Table 9, and the measured out-of-plane and in-plane flexural frequencies are shown in Tables 10 and 11. The Young's moduli and Poisson's ratios calculated from these results for the fundamental modes (n=1) using eqs. 6-8 are shown in Table 12. The Young's moduli for higher orders of vibration obtained using eq. 5 are shown in Tables S1-S6 in the SI. The shear moduli derived from the fundamental torsional frequencies were used for estimating Young's moduli and Poisson's ratios rather than those derived from the overtones because this value is more accurate. The method used for the glass bars was then used to determine the optimum shear constants of the novaculites. The comparison of the standard deviation of Young's modulus as a function of the shear constant are shown in Figures 10 and 11 for out-of-plane and in-plane flexural vibrations.

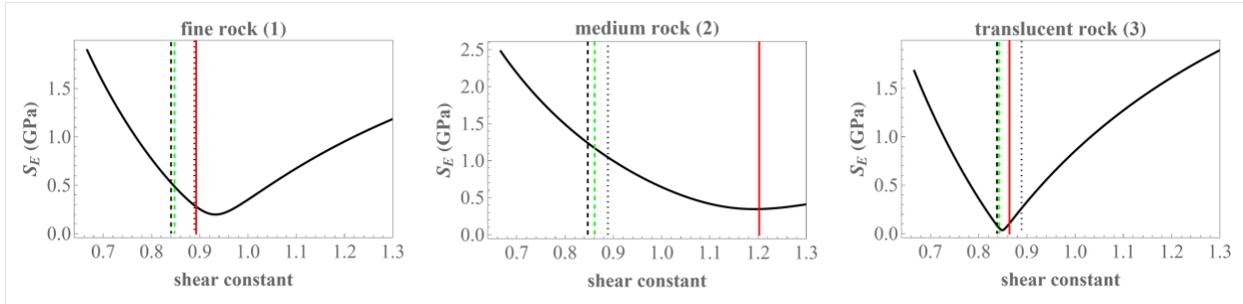

**Figure 10** Standard deviation of Young's modulus ($S_E$) as a function of the shear constant, showing the optimum shear constants of the novaculites (out-of-plane flexural vibrations, n=1-7). Picket (black dots), Cowper (black dashes), Kaneko (green dashes), Hutchinson (red line).

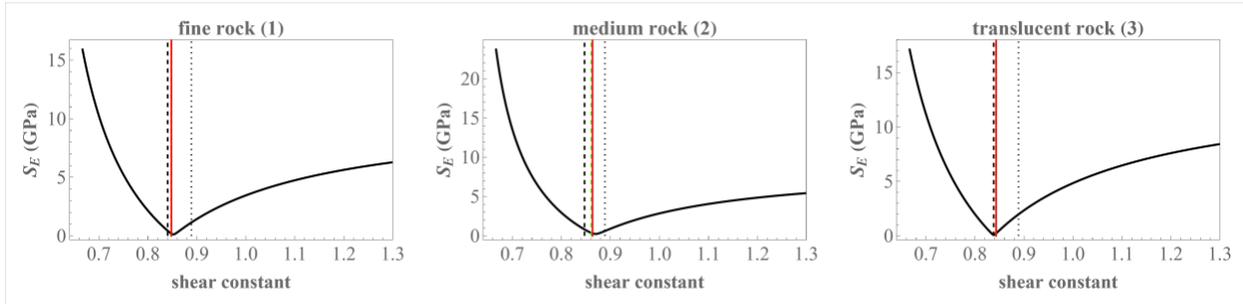

**Figure 11** Standard deviation of Young's modulus ($S_E$) as a function of the shear constant, showing the optimum shear constants of the novaculites (in-plane flexural vibrations, n=1-6). Picket (black dots), Cowper (black dashes), Kaneko (green dashes), Hutchinson (red line).

As expected from the glass bar results, the shear constants calculated using the method of Hutchinson closely matched the optimum shear constants for the novaculites for both in-plane and out-of-plane flexural modes of vibration. Kaneko's shear constant gives similar results for the in-plane flexural mode of vibration and cannot be easily distinguished from Hutchinson's (Figure 11). The estimated Young's moduli and Poisson's ratios of the novaculites are shown in Tables S1-S6. Poisson's ratio was calculated from the values of E and G (n=1) using eqs. 5 and 8 and the same iterative method employed for the glass bars.

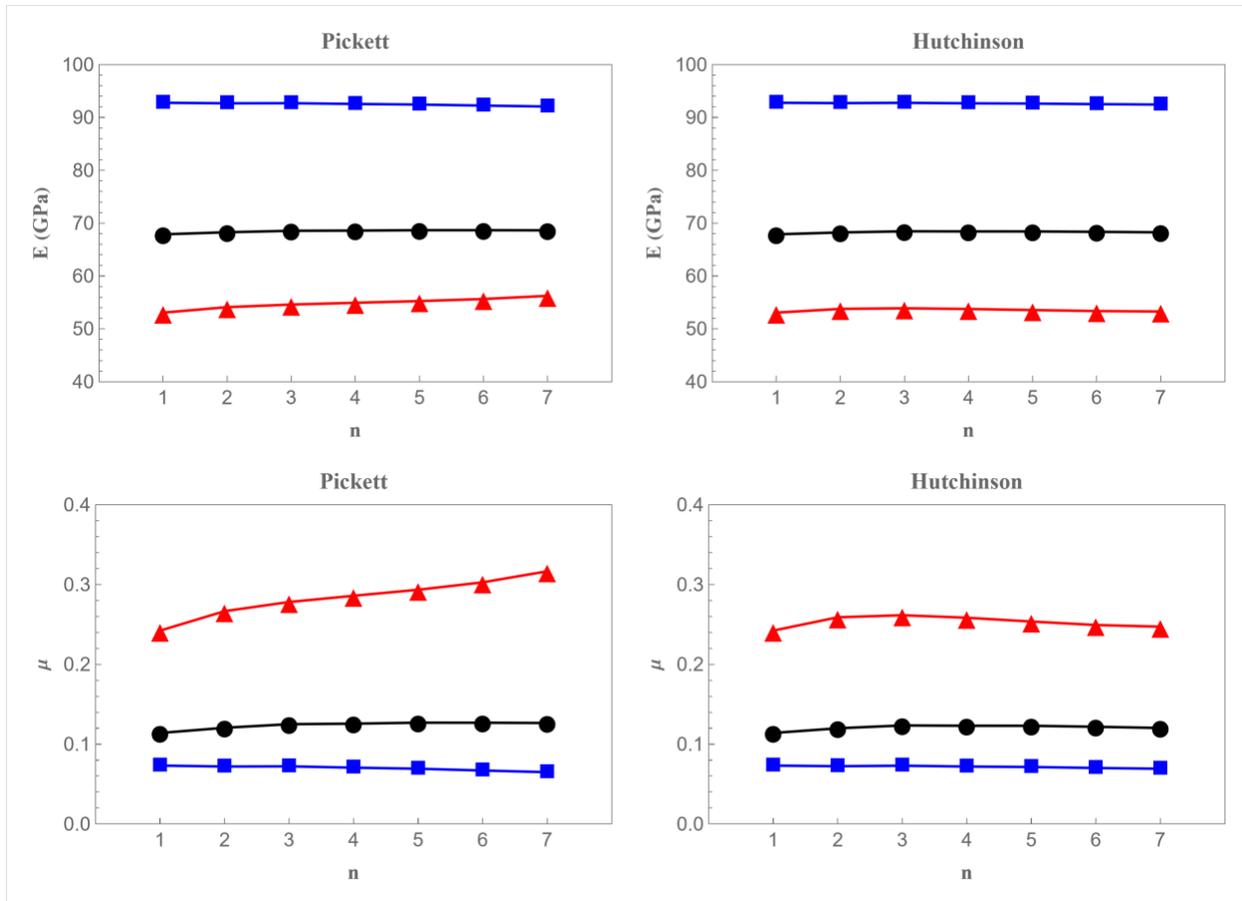

**Figure 12** Young's moduli and Poisson's ratios of novaculites (out-of-plane flexural mode). Fine rock (1) (black circles and line), Medium rock (2) (blue squares and line), Translucent rock (3) (red triangles and line).

These results (Tables S1-S6) show that the Hutchinson shear constant gives the most consistent results for the Young's moduli and Poisson's ratios (Figures 12 and 13 and Figures S4-S7). Using this shear constant, Poisson's ratio does not change significantly with increasing order of vibration for either out-of-plane or in-plane flexural vibration modes. However, using the Pickett shear constant, Poisson's ratio became negative for n > 6 (in-plane flexural mode) for the translucent novaculite (Figure 13). This approach also generated the largest differences between the out-of-plane and in-plane Young's moduli (Figure 14). The consistent Young's moduli and Poisson's ratios obtained under both out-of-plane and in-plane flexural modes of vibration using the Hutchinson shear constant imply that use of this shear constant is not necessarily limited to samples with b/t ≤ 2.4 provided the b/l ratio is ≤ ~0.17 (Figure 6).

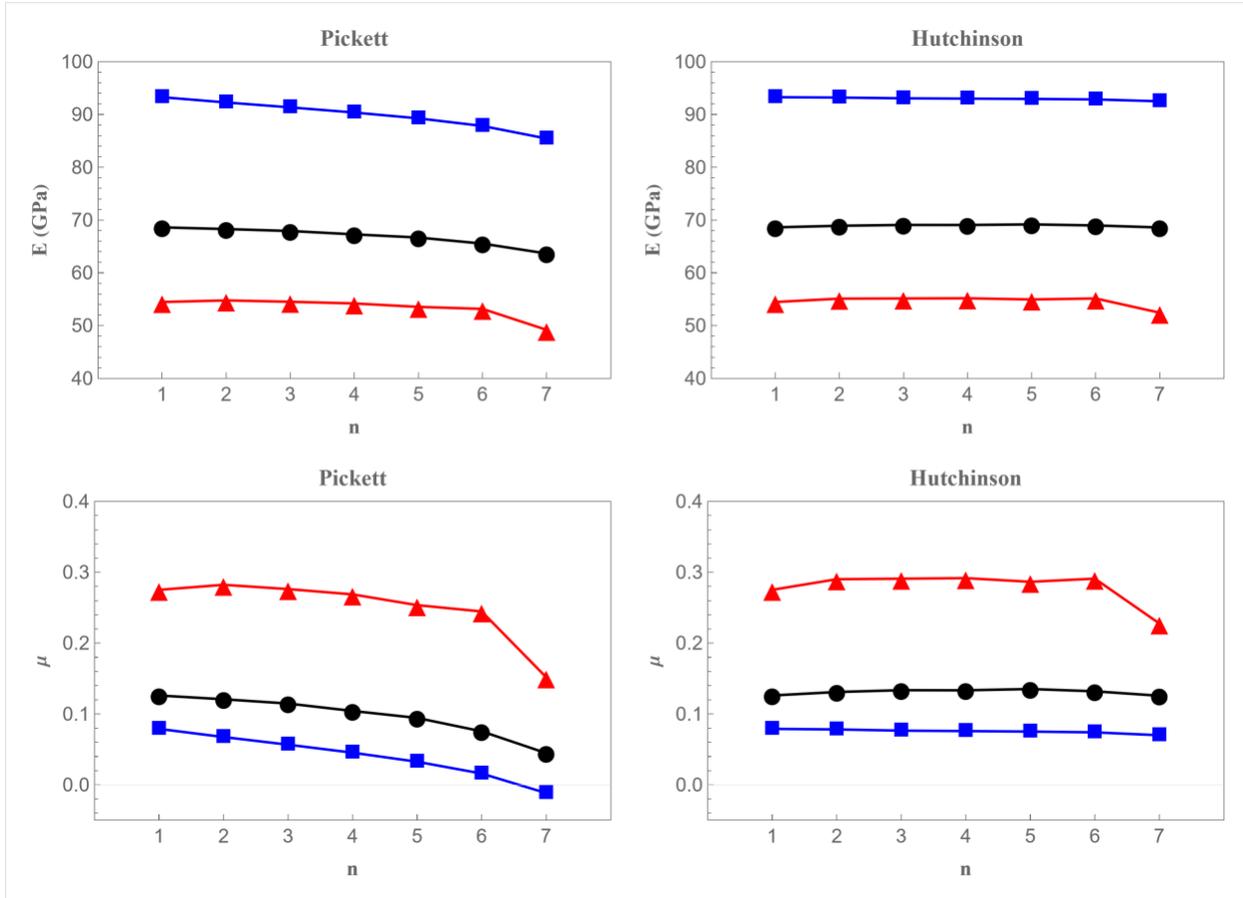

**Figure 13** Young's moduli and Poisson's ratios of novaculites (in-plane flexural mode). Fine rock (1) (black circles and line), Medium rock (2) (blue squares and line), Translucent rock (3) (red triangles and line).

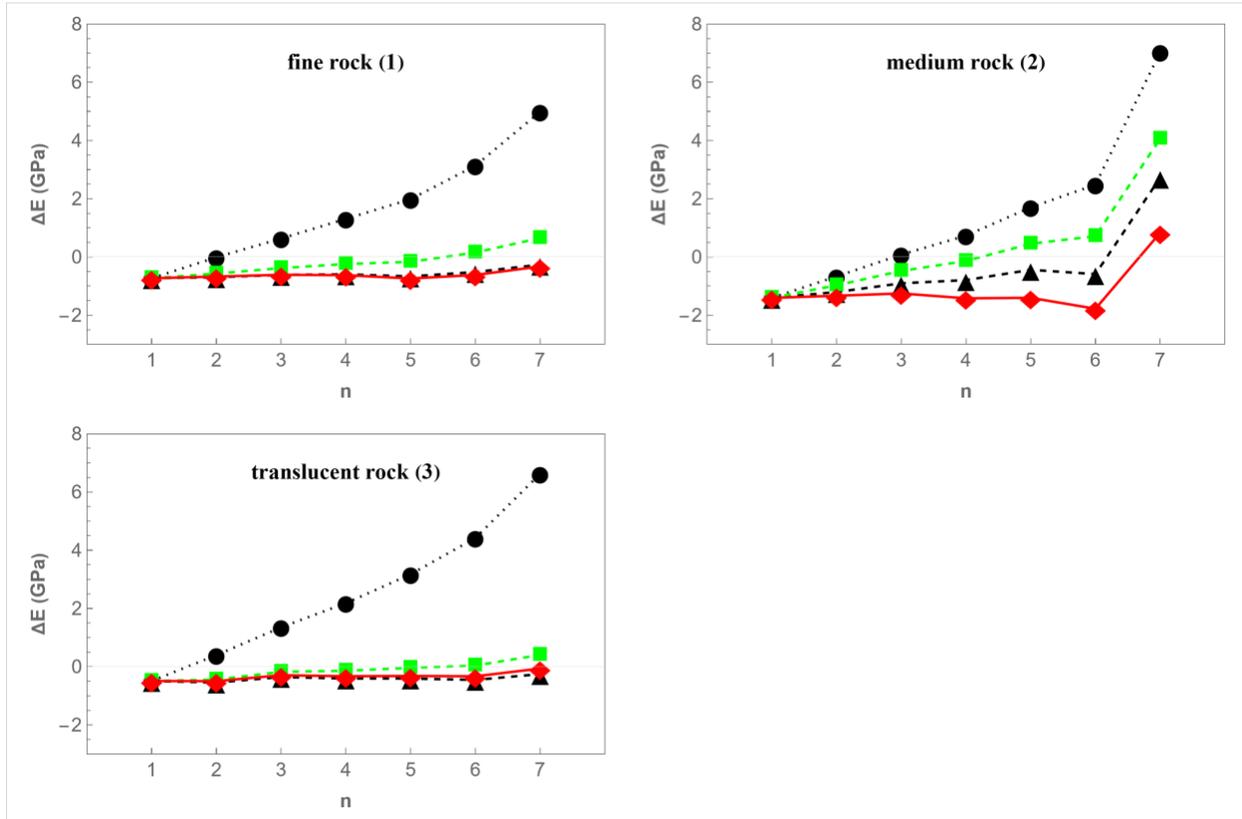

**Figure 14** The difference in Young's moduli from out-of-plane and in-plane flexural modes of vibration. Pickett (black circles and dots), Cowper (green squares and dashes), Kaneko (black triangles and dashes), Hutchinson (red diamonds and lines).

## 4    Conclusions

The choice of shear constant has a significant effect on the precision of the Young's modulus calculation for Timoshenko beams of rectangular cross-section, particularly at higher orders of vibration. Several shear constants have been proposed with the most recent being that of Hutchinson [14]. Our analysis shows that this shear constant works well for glass bars with width/thickness (b/t) ratios $\leq 2.4$, and that this useful ratio can be extended for relatively long glass samples with width/length (b/l) ratios $\leq 0.1$. When tested with porous materials, the Poisson's ratio did not change significantly with increasing order of vibration for both the out-of-plane and in-plane flexural modes using Hutchinson's shear constant. Hutchinson's shear constant yielded consistent elastic constants even though the b/t ratios were ~3.8. However, these samples had b/l ratios of ~0.17. Thus, the glass bar and rock results suggest that Hutchinson's shear constant can be used for rectangular samples with b/t ratios at least up to 2.4, and that this can be extended to larger values if the b/l ratio is ≤ 0.17. For all other circumstances, however, Kaneko's shear constant should be used. By selecting the shear constant appropriate to the b/t and b/l ratios, consistent Young's moduli can be obtained for orders of flexural vibration up to at least 7 using the impulse excitation (IE) technique.

## Acknowledgments

This material is based upon work supported by the U.S. Department of Energy, Office of Science, Office of Basic Energy Sciences, Geosciences program. The authors declare no competing financial interests.

## Tables

**Table 1** Shear constants from different sources.

| Source | k' |
|---|---|
| Timoshenko [8] | 2/3 |
| Pickett [12] | 8/9 |
| Ritchie [15] | 5/6 |
| Cowper [16] | $\dfrac{10 + 10\mu}{12 + 11\mu}$ |
| Kaneko [9] | $\dfrac{5 + 5\mu}{6 + 5\mu}$ |
| Hutchinson [14] | $-\dfrac{2(1 + \mu)}{\left[\dfrac{144}{t^5}C_4 + \mu\left(1 - \dfrac{b^2}{t^2}\right)\right]}$ |

$$C_4 = \frac{1}{720} t^3 b(-12t^2 - 15\mu t^2 + 5\mu b^2) + \sum_{n=1}^{\infty} \frac{\mu^2 b^5 \left(n\pi t - b \, tanh\left(\frac{n\pi t}{b}\right)\right)}{4(n\pi)^5 \, (1+\mu)}$$

b = width
t = thickness

**Table 2** Physical properties of the soda-lime silica glass bars.

| Sample | Length (mm) | Width (mm) | Thickness (mm) | Mass (g) | Density (kg/m³) | Width/ Length | Width/ Thickness | Length/ Thickness |
|---|---|---|---|---|---|---|---|---|
| 1 | 63.44 | 6.65 | 5.61 | 5.74 | 2423.0 | 0.1049 | 1.1857 | 11.3043 |
| 2 | 95.41 | 9.63 | 5.62 | 12.63 | 2446.3 | 0.1009 | 1.7120 | 16.9708 |
| 3 | 47.80 | 9.61 | 5.62 | 6.31 | 2442.0 | 0.2011 | 1.7091 | 8.4993 |
| 4 | 63.44 | 12.83 | 5.62 | 11.24 | 2456.8 | 0.2022 | 2.2808 | 11.2822 |
| 5 | 126.95 | 12.80 | 5.62 | 22.41 | 2455.4 | 0.1009 | 2.2803 | 22.6091 |
| 6 | 52.94 | 15.94 | 5.61 | 11.61 | 2450.2 | 0.3012 | 2.8399 | 9.4300 |
| 7 | 79.32 | 15.96 | 5.62 | 17.45 | 2454.0 | 0.2012 | 2.8403 | 14.1189 |
| 8 | 158.74 | 16.01 | 5.62 | 35.17 | 2461.9 | 0.1008 | 2.8463 | 28.2305 |
| 9 | 47.66 | 19.14 | 5.62 | 12.62 | 2460.5 | 0.4016 | 3.4048 | 8.4774 |
| 10 | 63.59 | 19.13 | 5.62 | 16.81 | 2460.8 | 0.3008 | 3.4048 | 11.3210 |
| 11 | 95.14 | 19.18 | 5.62 | 25.32 | 2467.3 | 0.2016 | 3.4115 | 16.9198 |
| 12 | 190.53 | 19.15 | 5.62 | 50.58 | 2467.6 | 0.1005 | 3.4074 | 33.9082 |
| 13 | 74.17 | 22.34 | 5.62 | 22.99 | 2467.3 | 0.3012 | 3.9719 | 13.1881 |
| 14 | 111.18 | 22.32 | 5.62 | 34.38 | 2466.7 | 0.2007 | 3.9733 | 19.7935 |
| 15 | 222.20 | 22.34 | 5.62 | 68.94 | 2471.3 | 0.1006 | 3.9763 | 39.5444 |
| 16 | 63.75 | 25.47 | 5.63 | 22.57 | 2470.4 | 0.3996 | 4.5277 | 11.3313 |
| 17 | 84.62 | 25.48 | 5.63 | 29.98 | 2471.5 | 0.3012 | 4.5305 | 15.0436 |
| 18 | 126.86 | 25.46 | 5.62 | 44.83 | 2470.7 | 0.2007 | 4.5334 | 22.5850 |
| 19 | 253.72 | 25.47 | 5.62 | 89.90 | 2475.1 | 0.1004 | 4.5326 | 45.1459 |

**Table 3** Physical properties of the novaculite samples.

| Sample | Length (mm) | Width (mm) | Thickness (mm) | Mass (g) | Density (kg/m³) | Width/ Length | Width/ Thickness | Length/ Thickness |
|---|---|---|---|---|---|---|---|---|
| fine rock (1) | 151.36 | 25.32 | 6.985 | 63.64 | 2377.0 | 0.1673 | 3.6253 | 21.6693 |
| medium rock (2) | 152.30 | 25.66 | 6.664 | 58.25 | 2236.3 | 0.1685 | 3.8511 | 22.8541 |
| translucent rock (3) | 152.71 | 25.43 | 6.730 | 68.62 | 2626.1 | 0.1665 | 3.7779 | 22.6909 |

**Table 4** Out-of-plane flexural frequencies (Hz) of glass bars for n = 1-7.

| Sample | 1 | 2 | 3 | 4 | 5 | 6 | 7 |
|---|---|---|---|---|---|---|---|
| 1 | 7526 | 19853 | 36809 | 57027 | 79562 | 103621 | 128649 |
| 2 | 3382 | 9129 | 17407 | 27782 | 39903 | 53388 | 67962 |
| 3 | 13064 | 33540 | 60395 | 90999 | 123864 | 157656 | 191470 |

| 4 | 7556 | 19952 | 36997 | 57326 | 79951 | 103987 | 128378 |
| 5 | 1919 | 5226 | 10080 | 16321 | 23788 | 32298 | 41725 |
| 6 | 10732 | 27906 | 50825 | 77283 | 106024 | 137824 | 168633 |
| 7 | 4869 | 13106 | 24637 | 38888 | 55177 | 72919 | 91310 |
| 8 | 1231 | 3371 | 6537 | 10661 | 15699 | 21465 | 27981 |
| 9 | 13169 | 33844 | 60633 | 92546 | 125423 | 158860 | 194890 |
| 10 | 7515 | 19874 | 36848 | 57008 | 79328 | 104676 | 129636 |
| 11 | 3408 | 9217 | 17584 | 28085 | 40321 | 53893 | 68100 |
| 12 | 854.8 | 2343 | 4563 | 7476 | 11042 | 15215 | 19951 |
| 13 | 5575 | 14908 | 27993 | 43869 | 61704 | 82343 | 102970 |
| 14 | 2500 | 6796 | 13063 | 21044 | 30479 | 41102 | 52415 |
| 15 | 629.4 | 1731 | 3372 | 5538 | 8206 | 11351 | 14943 |
| 16 | 7492 | 19830 | 36568 | 57507 | 79937 | 103698 | 129398 |
| 17 | 4303 | 11590 | 21957 | 34748 | 49306 | 66405 | 83747 |
| 18 | 1926 | 5255 | 10146 | 16443 | 23966 | 32523 | 41715 |
| 19 | 482.8 | 1328 | 2594 | 4265 | 6335 | 8787 | 11598 |

**Table 5** Young's modulus (E)[1], shear modulus (G), and Poisson's ratio ($\mu$) for the glass bars (n=1).

| Sample | E (GPa) | G (GPa) | $\mu$ |
|---|---|---|---|
| 1 | 70.36 | 29.35 | 0.199 |
| 2 | 71.08 | 29.80 | 0.193 |
| 3 | 71.19 | 29.83 | 0.193 |
| 4 | 71.65 | 29.98 | 0.195 |
| 5 | 71.44 | 29.99 | 0.191 |
| 6 | 71.63 | 30.04 | 0.193 |
| 7 | 71.42 | 30.05 | 0.188 |
| 8 | 71.51 | 30.00 | 0.192 |
| 9 | 72.15 | 29.87 | 0.208 |
| 10 | 71.79 | 30.02 | 0.196 |
| 11 | 71.96 | 30.01 | 0.199 |
| 12 | 71.65 | 30.03 | 0.193 |
| 13 | 72.17 | 30.02 | 0.202 |
| 14 | 71.90 | 30.07 | 0.196 |
| 15 | 71.85 | 30.02 | 0.197 |
| 16 | 72.13 | 29.83 | 0.209 |
| 17 | 72.31 | 29.99 | 0.206 |
| 18 | 72.16 | 30.05 | 0.201 |
| 19 | 71.88 | 30.00 | 0.198 |
| Mean | 71.70 | 29.94 | 0.197 |
| Std. dev.[2] | 0.47 | 0.17 | 0.006 |

[1]E derived from fundamental out-of-plane flexural frequencies with eqs. 6 and 7

**Table 6** RMSs of the shear constants (out-of-plane flexure).

| Source | RMS |
|---|---|
| **Pickett** | 0.0807 |
| **Kaneko** | 0.0941 |
| **Cowper** | 0.1012 |
| **Ritchie** | 0.1090 |
| **Hutchinson** | 0.1746 |
| **Timoshenko** | 0.2549 |

**Table 7** In-plane flexural frequencies (Hz) of glass bars for n = 1-7.

| Sample | 1 | 2 | 3 | 4 | 5 | 6 | 7 |
|---|---|---|---|---|---|---|---|
| **1** | 8638 | 22508 | 41130 | 62882 | 86651 | / | / |
| **12** | 2798 | 7304 | 13381 | 20506 | 28314 | 36584 | 45086 |
| **18** | 7747 | 18267 | 30595 | 43349 | 55918 | / | / |
| **19** | 2105 | 5495 | 10065 | 15422 | 21299 | 27493 | 33881 |

**Table 8** RMSs of the shear constants (in-plane flexure).

| Source | RMS |
|---|---|
| **Kaneko** | 0.0119 |
| **Hutchinson** | 0.0139 |
| **Cowper** | 0.0190 |
| **Ritchie** | 0.0292 |
| **Pickett** | 0.0309 |
| **Timoshenko** | 0.1938 |

**Table 9** Shear modulus (GPa) of the novaculites for n = 1-7.

| Sample | 1 | 2 | 3 | 4 | 5 | 6 | 7 | Mean | Std. dev. |
|---|---|---|---|---|---|---|---|---|---|
| **fine rock (1)** | 30.47 | 30.64 | 30.53 | 30.38 | 30.30 | 30.27 | 30.18 | 30.39 | 0.16 |
| **medium rock (2)** | 21.36 | 21.92 | 21.98 | 21.95 | 22.10 | 22.23 | 22.66 | 22.03 | 0.39 |
| **translucent rock (3)** | 43.22 | 43.02 | 42.70 | 42.34 | 42.08 | 41.94 | 41.58 | 42.41 | 0.60 |

**Table 10** Out-of-plane flexural frequencies (Hz) of the novaculites for n = 1-7.

| Sample | 1 | 2 | 3 | 4 | 5 | 6 | 7 |
|---|---|---|---|---|---|---|---|
| **fine rock (1)** | 1663 | 4541 | 8766 | 14180 | 20656 | 28035 | 36184 |
| **medium rock (2)** | 1429 | 3930 | 7608 | 12343 | 18023 | 24541 | 31817 |
| **translucent rock (3)** | 1752 | 4774 | 9216 | 14927 | 21777 | 29611 | 38300 |

**Table 11** In-plane flexural frequencies (Hz) of the novaculites for n = 1-7.

| Sample | 1 | 2 | 3 | 4 | 5 | 6 | 7 |
|---|---|---|---|---|---|---|---|
| **fine rock (1)** | 5615 | 13741 | 23705 | 34431 | 45457 | 56274 | 66459 |
| **medium rock (2)** | 5135 | 12522 | 21430 | 30932 | 40541 | 49988 | 58119 |
| **translucent rock (3)** | 6153 | 15071 | 26034 | 37910 | 50134 | 62248 | 73747 |

Note: The in-plane flexural frequency for the medium rock (2) at n = 7 was difficult to excite and deemed unreliable.

**Table 12** Young modulus (GPa) and Poisson's ratio of the novaculites (n=1).

| Sample | $E_{out-of-plane}$ | $E_{in-plane}$ | $\mu_{out-of-plane}$ | $\mu_{in-plane}$ |
|---|---|---|---|---|
| **fine rock (1)** | 67.88 | 68.61 | 0.114 | 0.126 |
| **medium rock (2)** | 53.06 | 54.47 | 0.242 | 0.275 |
| **translucent rock (3)** | 92.78 | 93.27 | 0.073 | 0.079 |

Note: The flexural mode of vibration is indicated by the subscripts.

## Appendix: Example Calculations for Sample 8

**General information of sample 8**

Physical properties: Length (l) = 158.74 mm, width (b) = 16.01 mm, thickness (t) = 5.62 mm, density ($\rho$) = 2461.9 kg/m$^3$

Out-of-plane resonant frequencies (n = 1-7): 1231 Hz, 3371 Hz, 6537Hz, 10661 Hz, 15699 Hz, 21465 Hz, 27981 Hz

Shear modulus: G = 30.00 GPa

**Calculating optimum shear constant, $k'_{opt}$**

By assuming various values of shear constants, k', the only unknown in eq. 2 was E. Eq. 2 was then solved for E using a numerical function, *FindRoot*, built into Mathematica (version 13.0.1.0). The following procedure was used:

1. Calculate r

2. Set k' = 0.67 (2/3 as per Timoshenko)

3. Calculate $\omega$ using the fundamental out-of-plane flexural frequency (n=1)

4. Use *FindRoot* to solve for E in eq. 2

5. Repeat steps 3 and 4 stepping through increasing orders of resonant frequencies

6. Take the sample standard deviation for each of the seven Young's moduli

7. Repeat steps 2-6 incrementing k' by 0.01

8. Stop when k' = 1.3

9. The optimum k' is that giving the minimum standard deviation

Note: *FindRoot* requires an initial guess for the value being solved. For n = 1, a value of 71 GPa was used (the value of 71.51 GPa given in Table 5, rounded down). For higher values of n, the previous value of E was used as the guess. A plot of the sample standard deviation vs shear constant is shown in Figure A1.

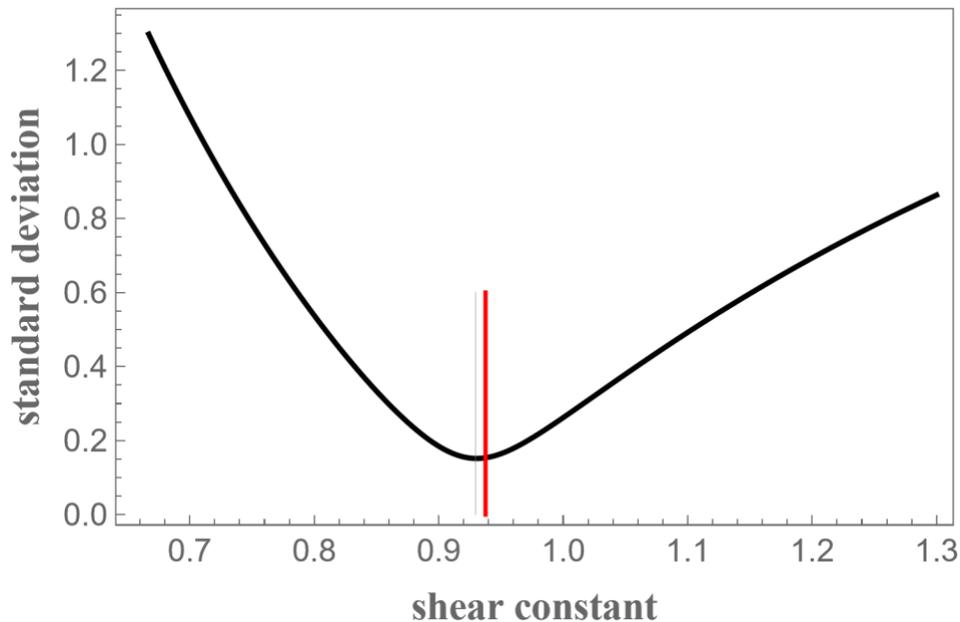

**Figure A1** standard deviation vs shear constant for sample 8. Gray line indicates optimum shear constant (0.9297), red line is the Hutchinson shear constant (0.9378).

The Hutchison shear constant shown in the figure was calculated from the formula shown in Table 1, using the values of b and t listed above and a Poisson's ratio, μ, of 0.192 (Table 3).

Note: the summation was carried out for the first 10000 values of n. Thus, the difference between Hutchinson's shear constant and k' $(\Delta k')$ = 0.9379 - 0.9307 = 0.0072.

**Calculating Young's modulus using shear constants depending on the Poisson's ratio**

We can calculate E as before but this time we use the values of k' given in Table 1 rather than sweep through a range. However, a complication arises for the shear constants of Cowper, Kaneko, and Hutchinson as these also rely on the Poison's ratio, μ. Fortunately, we can use eq. 8. So, once again, the only unknown is E.

# Supporting Information

Improved Calculation of the Young's Modulus of Rectangular Prisms from their
Resonant Frequency Overtones by Identifying Appropriate Shear Constants

Paul A. Bosomworth[1], Rui Zhang[2, &], Lawrence M. Anovitz[2, *]

[1]BuzzMac International LLC, Portland, ME 04101, United States
[2]Chemical Sciences Division, Oak Ridge National Laboratory, Oak Ridge, TN 37831, United States

[*]Corresponding author: Lawrence M. Anovitz, anovitzlm@ornl.gov
[&]Present address: Eyring Materials Center, Arizona State University, Tempe, AZ 85287, United States

**Figures**

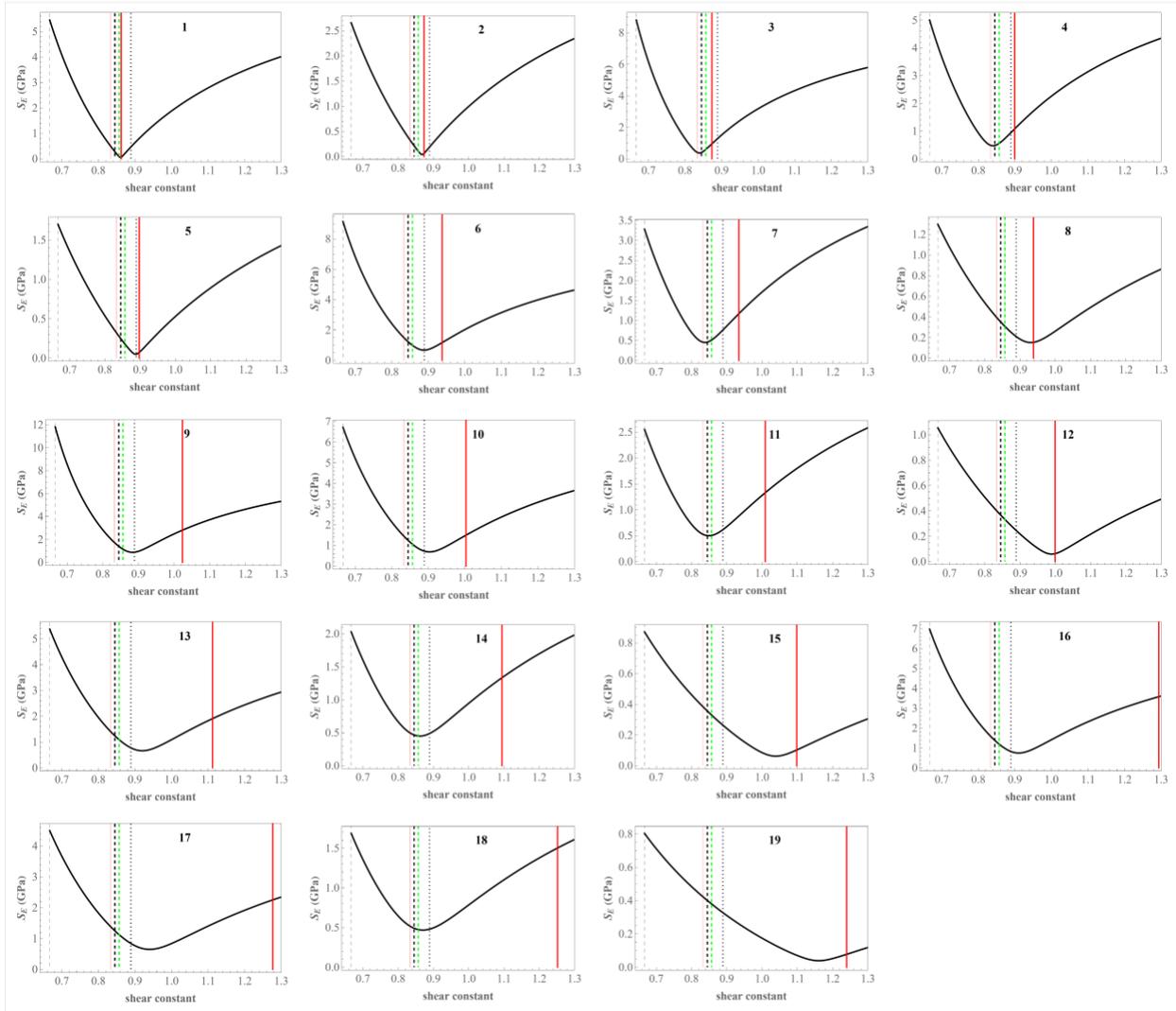

**Figure S1** Standard deviation of Young's modulus ($S_E$) as a function of the shear constant, showing the optimum shear constants for all the glass bars (out-of-plane flexural vibrations, n=1-7). Timoshenko (thin gray dashes), Ritchie (thin red line), Picket (black dots), Cowper (black dashes), Kaneko (green dashes), Hutchinson (thick red line).

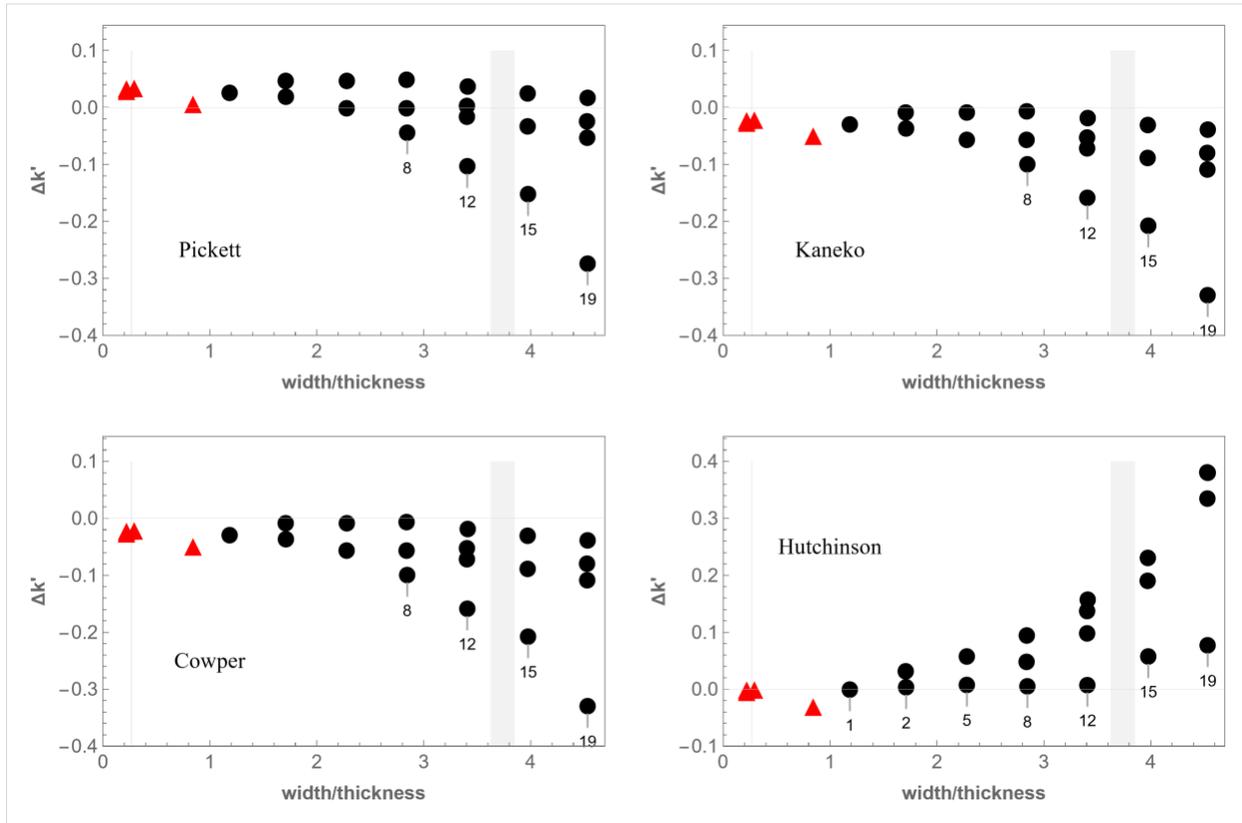

**Figure S2** Difference between shear constant and the optimum ($\Delta k'$) for each literature shear constant. Out-of-plane flexure mode (black circles), in-plane flexure mode (red triangles).

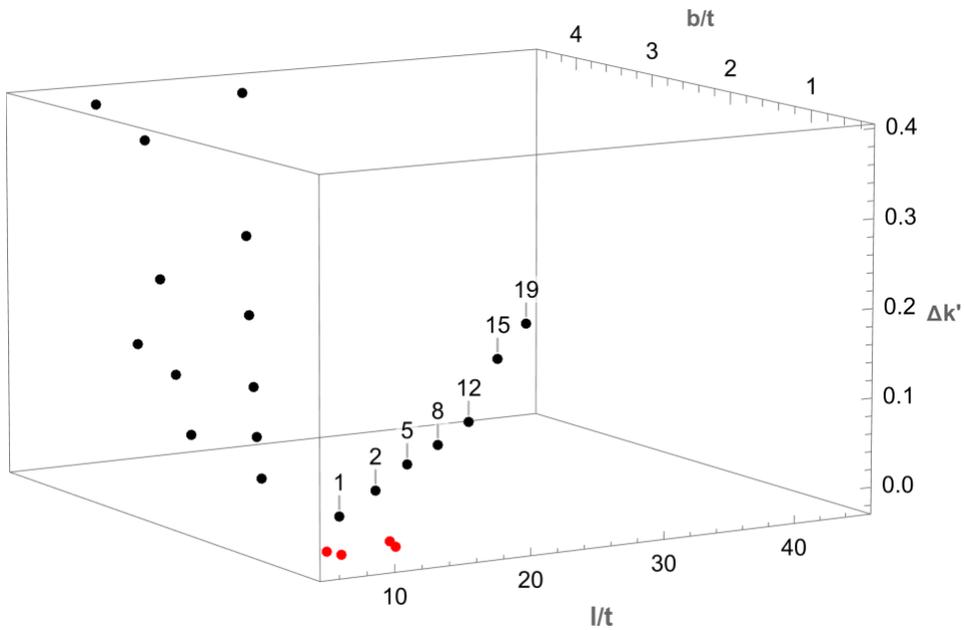

**Figure S3** Difference between Hutchinson's shear constant and the optimum ($\Delta k'$). Out-of-plane flexure mode (black), in-plane flexure mode (red).

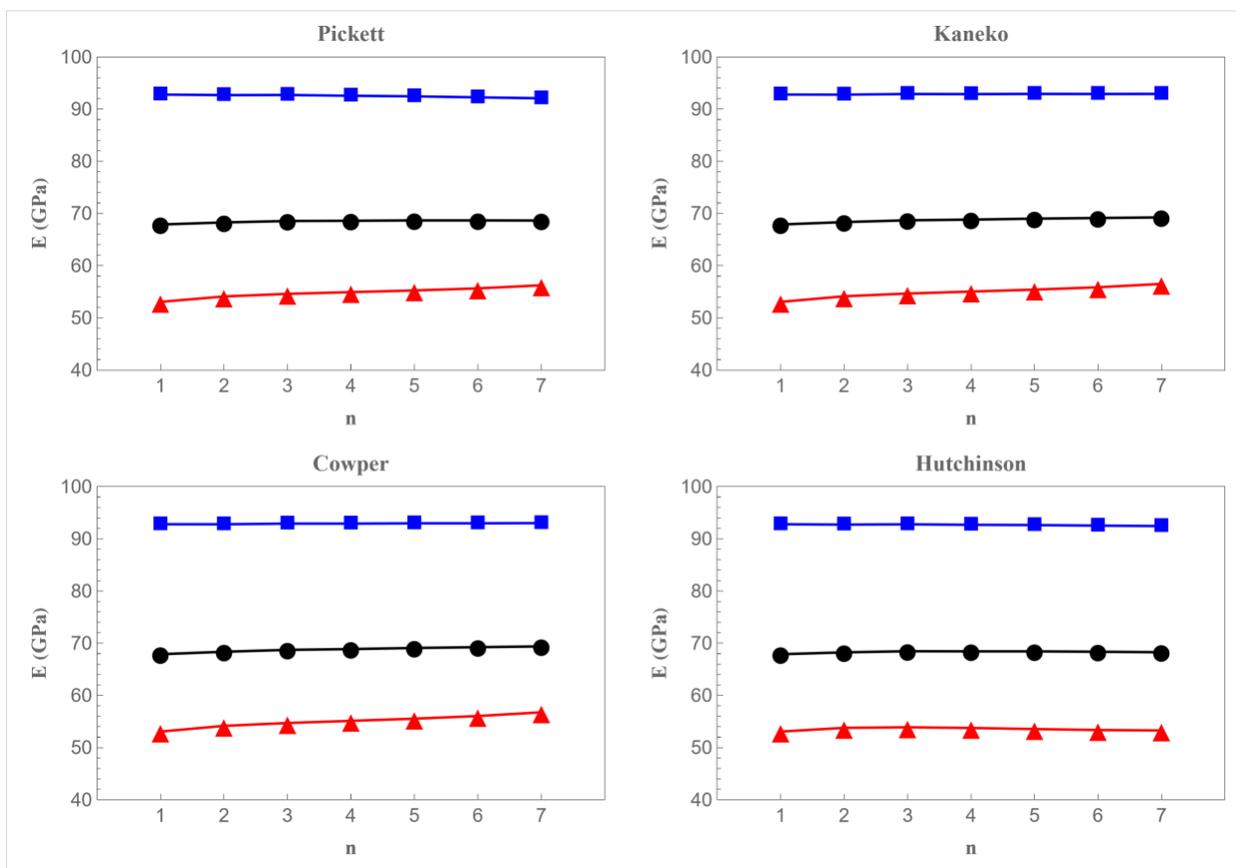

**Figure S4** Young's moduli of the novaculites (out-of-plane flexural mode).

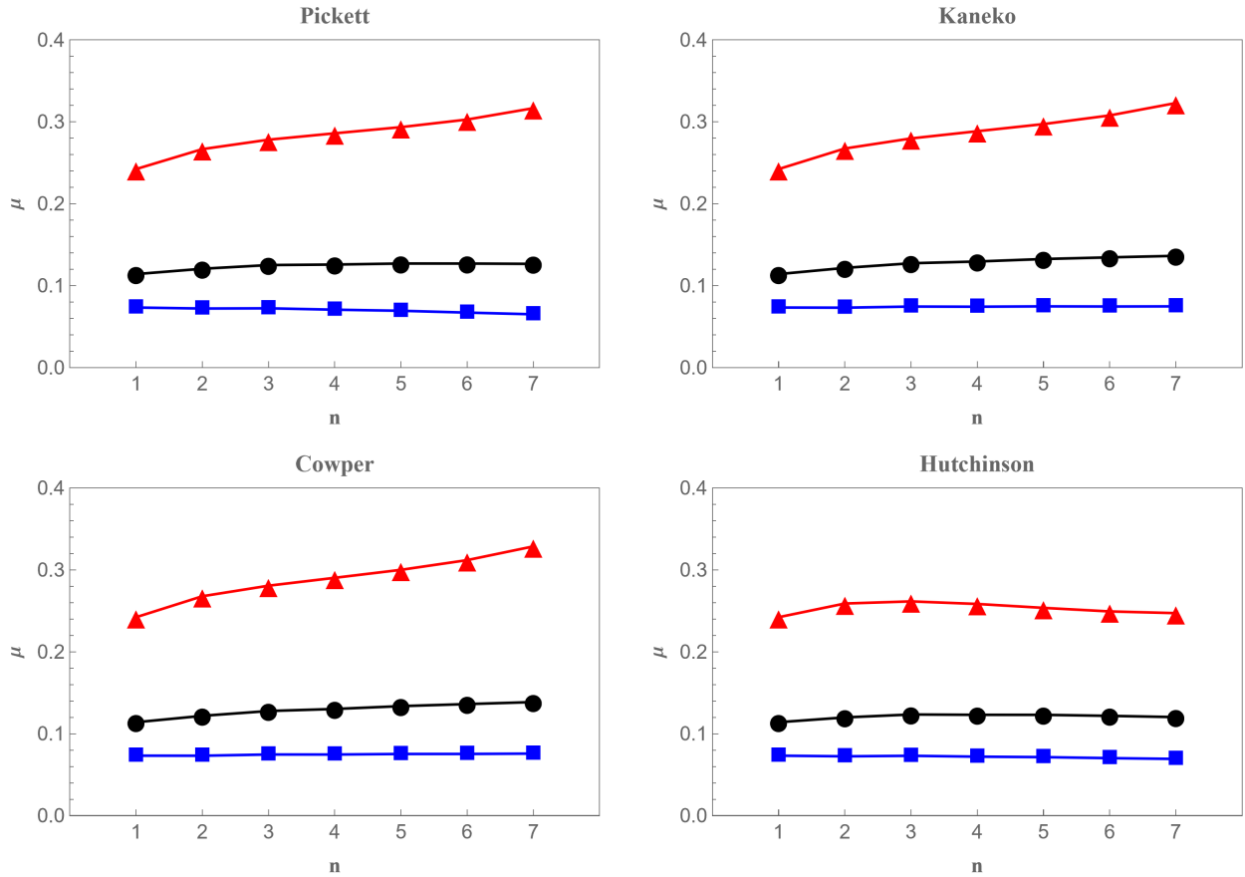

**Figure S5** Poisson's ratios of the novaculites (out-of-plane flexural mode).

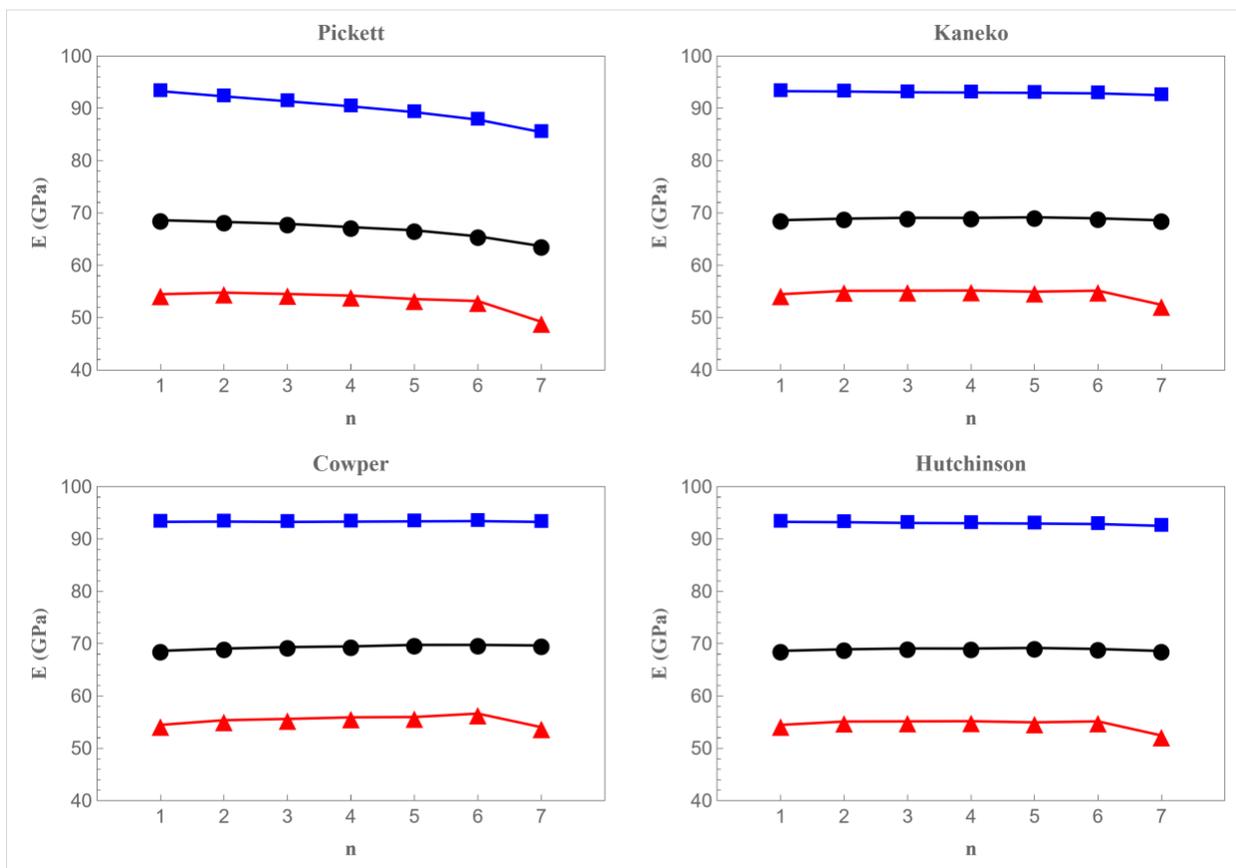

**Figure S6** Young's moduli of the novaculites (in-plane flexural mode).

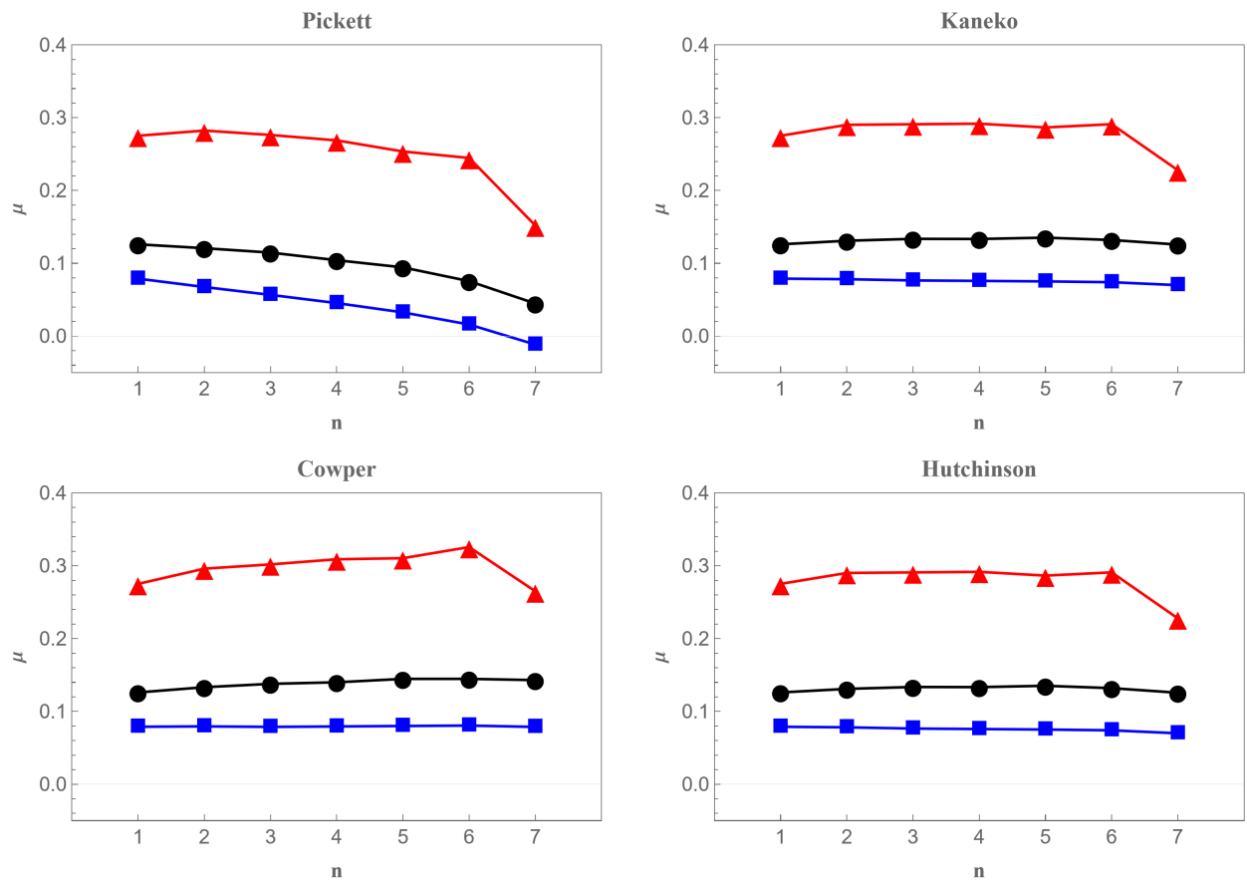

**Figure S7** Poisson's ratios of the novaculites (in-plane flexural mode).

**Tables**

**Table S1** Young's modulus (GPa) and Poisson's ratios (out-of-plane) of the fine rock (1).

| n | $E_{Pickett}$ | $E_{Cowper}$ | $E_{Kaneko}$ | $E_{Hutchinson}$ | $\mu_{Pickett}$ | $\mu_{Cowper}$ | $\mu_{Kaneko}$ | $\mu_{Hutchinson}$ |
|---|---|---|---|---|---|---|---|---|
| 1 | 67.88 | 67.88 | 67.88 | 67.88 | 0.114 | 0.114 | 0.114 | 0.114 |
| 2 | 68.28 | 68.36 | 68.34 | 68.24 | 0.121 | 0.122 | 0.122 | 0.120 |
| 3 | 68.55 | 68.72 | 68.69 | 68.46 | 0.125 | 0.128 | 0.127 | 0.123 |
| 4 | 68.59 | 68.87 | 68.82 | 68.43 | 0.126 | 0.130 | 0.129 | 0.123 |
| 5 | 68.67 | 69.08 | 69.00 | 68.43 | 0.127 | 0.134 | 0.132 | 0.123 |
| 6 | 68.67 | 69.23 | 69.13 | 68.36 | 0.127 | 0.136 | 0.134 | 0.122 |
| 7 | 68.64 | 69.38 | 69.24 | 68.26 | 0.127 | 0.139 | 0.136 | 0.120 |
| Mean | 68.47 | 68.79 | 68.73 | 68.30 | 0.124 | 0.129 | 0.128 | 0.121 |
| Std. dev. | 0.29 | 0.52 | 0.48 | 0.20 | 0.005 | 0.009 | 0.008 | 0.003 |

**Table S2** Young's modulus (GPa) and Poisson's ratios (out-of-plane) of the medium rock (2).

| n | $E_{Pickett}$ | $E_{Cowper}$ | $E_{Kaneko}$ | $E_{Hutchinson}$ | $\mu_{Pickett}$ | $\mu_{Cowper}$ | $\mu_{Kaneko}$ | $\mu_{Hutchinson}$ |
|---|---|---|---|---|---|---|---|---|
| 1 | 53.06 | 53.06 | 53.06 | 53.06 | 0.242 | 0.242 | 0.242 | 0.242 |
| 2 | 54.10 | 54.16 | 54.14 | 53.78 | 0.267 | 0.268 | 0.267 | 0.259 |
| 3 | 54.60 | 54.71 | 54.66 | 53.89 | 0.278 | 0.281 | 0.280 | 0.262 |
| 4 | 54.93 | 55.12 | 55.04 | 53.75 | 0.286 | 0.290 | 0.288 | 0.258 |
| 5 | 55.25 | 55.53 | 55.41 | 53.55 | 0.293 | 0.300 | 0.297 | 0.254 |
| 6 | 55.65 | 56.04 | 55.86 | 53.36 | 0.303 | 0.312 | 0.308 | 0.249 |
| 7 | 56.24 | 56.76 | 56.51 | 53.28 | 0.317 | 0.329 | 0.323 | 0.247 |
| Mean | 54.83 | 55.05 | 54.95 | 53.52 | 0.284 | 0.289 | 0.287 | 0.253 |
| Std. dev. | 1.04 | 1.22 | 1.14 | 0.30 | 0.024 | 0.029 | 0.027 | 0.007 |

**Table S3** Young's modulus (GPa) and Poisson's ratios (out-of-plane) of the translucent rock (3).

| n | $E_{Pickett}$ | $E_{Cowper}$ | $E_{Kaneko}$ | $E_{Hutchinson}$ | $\mu_{Pickett}$ | $\mu_{Cowper}$ | $\mu_{Kaneko}$ | $\mu_{Hutchinson}$ |
|---|---|---|---|---|---|---|---|---|
| 1 | 92.78 | 92.78 | 92.78 | 92.78 | 0.073 | 0.073 | 0.073 | 0.073 |
| 2 | 92.68 | 92.77 | 92.76 | 92.71 | 0.072 | 0.073 | 0.073 | 0.072 |
| 3 | 92.70 | 92.90 | 92.88 | 92.77 | 0.072 | 0.075 | 0.074 | 0.073 |
| 4 | 92.55 | 92.90 | 92.86 | 92.67 | 0.071 | 0.075 | 0.074 | 0.072 |
| 5 | 92.44 | 92.96 | 92.90 | 92.63 | 0.069 | 0.075 | 0.075 | 0.071 |
| 6 | 92.25 | 92.96 | 92.89 | 92.52 | 0.067 | 0.075 | 0.074 | 0.070 |
| 7 | 92.07 | 93.00 | 92.90 | 92.44 | 0.065 | 0.076 | 0.075 | 0.069 |
| Mean | 92.49 | 92.90 | 92.86 | 92.64 | 0.070 | 0.075 | 0.074 | 0.072 |
| Std. dev. | 0.26 | 0.09 | 0.06 | 0.13 | 0.003 | 0.001 | 0.001 | 0.001 |

**Table S4** Young's modulus (GPa) and Poisson's ratios (in-plane) of the fine rock (1).

| n | $E_{Pickett}$ | $E_{Cowper}$ | $E_{Kaneko}$ | $E_{Hutchinson}$ | $\mu_{Pickett}$ | $\mu_{Cowpe}$ | $\mu_{Kaneko}$ | $\mu_{Hutchinson}$ |
|---|---|---|---|---|---|---|---|---|
| 1 | 68.61 | 68.61 | 68.61 | 68.61 | 0.126 | 0.126 | 0.126 | 0.126 |
| 2 | 68.29 | 69.05 | 68.91 | 68.91 | 0.121 | 0.133 | 0.131 | 0.131 |
| 3 | 67.92 | 69.34 | 69.07 | 69.07 | 0.115 | 0.138 | 0.134 | 0.134 |
| 4 | 67.29 | 69.47 | 69.06 | 69.06 | 0.104 | 0.140 | 0.133 | 0.133 |
| 5 | 66.69 | 69.75 | 69.17 | 69.17 | 0.094 | 0.145 | 0.135 | 0.135 |
| 6 | 65.53 | 69.75 | 68.98 | 68.98 | 0.075 | 0.145 | 0.132 | 0.132 |
| 7 | 63.66 | 69.65 | 68.59 | 68.59 | 0.045 | 0.143 | 0.126 | 0.126 |
| Mean | 66.86 | 69.37 | 68.92 | 68.91 | 0.097 | 0.139 | 0.131 | 0.131 |
| Std. dev. | 1.75 | 0.42 | 0.23 | 0.23 | 0.029 | 0.007 | 0.004 | 0.004 |

**Table S5** Young's modulus (GPa) and Poisson's ratios (in-plane) of the medium rock (2).

| n | $E_{Pickett}$ | $E_{Cowper}$ | $E_{Kaneko}$ | $E_{Hutchinson}$ | $\mu_{Pickett}$ | $\mu_{Cowper}$ | $\mu_{Kaneko}$ | $\mu_{Hutchinson}$ |
|---|---|---|---|---|---|---|---|---|
| 1 | 54.47 | 54.47 | 54.47 | 54.47 | 0.275 | 0.079 | 0.079 | 0.079 |
| 2 | 54.77 | 55.36 | 55.11 | 55.11 | 0.282 | 0.079 | 0.078 | 0.078 |
| 3 | 54.52 | 55.61 | 55.14 | 55.14 | 0.276 | 0.079 | 0.076 | 0.076 |
| 4 | 54.20 | 55.91 | 55.18 | 55.18 | 0.269 | 0.079 | 0.076 | 0.076 |
| 5 | 53.55 | 55.98 | 54.96 | 54.95 | 0.254 | 0.080 | 0.075 | 0.075 |
| 6 | 53.17 | 56.63 | 55.15 | 55.14 | 0.245 | 0.081 | 0.074 | 0.074 |
| Mean | 54.11 | 55.66 | 55.00 | 55.00 | 0.267 | 0.080 | 0.076 | 0.076 |
| Std. dev. | 0.62 | 0.72 | 0.27 | 0.27 | 0.015 | 0.001 | 0.002 | 0.002 |

Note: The 7[th] order vibration (n = 7) could not be measured reliably for this novaculite.

**Table S6** Young's modulus (GPa) and Poisson's ratios (in-plane) of the translucent rock (3).

| n | $E_{Pickett}$ | $E_{Cowper}$ | $E_{Kaneko}$ | $E_{Hutchinson}$ | $\mu_{Pickett}$ | $\mu_{Cowper}$ | $\mu_{Kaneko}$ | $\mu_{Hutchinson}$ |
|---|---|---|---|---|---|---|---|---|
| 1 | 93.27 | 93.27 | 93.27 | 93.27 | 0.079 | 0.079 | 0.079 | 0.079 |
| 2 | 92.28 | 93.32 | 93.21 | 93.21 | 0.067 | 0.079 | 0.078 | 0.078 |
| 3 | 91.35 | 93.26 | 93.06 | 93.06 | 0.057 | 0.079 | 0.076 | 0.076 |
| 4 | 90.37 | 93.31 | 93.01 | 93.00 | 0.045 | 0.079 | 0.076 | 0.076 |
| 5 | 89.27 | 93.37 | 92.95 | 92.95 | 0.033 | 0.080 | 0.075 | 0.075 |
| 6 | 87.83 | 93.42 | 92.85 | 92.85 | 0.016 | 0.081 | 0.074 | 0.074 |
| 7 | 85.45 | 93.25 | 92.50 | 92.50 | -0.012 | 0.079 | 0.070 | 0.070 |
| Mean | 89.97 | 93.31 | 92.98 | 92.98 | 0.041 | 0.079 | 0.075 | 0.075 |
| Std. dev. | 2.70 | 0.06 | 0.26 | 0.26 | 0.031 | 0.001 | 0.003 | 0.003 |